\documentclass[a4paper,11pt]{article}
\usepackage{jheppub} 
\usepackage[utf8]{inputenc}
\usepackage{physics}
\usepackage{slashed}
\usepackage{caption}
\usepackage{xcolor}
\usepackage{tikz-feynman}
\usepackage{comment}
\usepackage{multirow}
\usepackage{graphics}
\usepackage{float}
\usepackage{cases}
\usepackage{cancel}
\usepackage{soul}
\usepackage{array}
\usepackage{mathtools}  
\usepackage{amsfonts}
\usepackage{hyperref}
\usepackage{amsmath}
\usepackage{amssymb}
\usepackage{tcolorbox}
\usepackage{tikz}

\title{Unitarity, Recursion and Soft Limits in (EA)dS through Dressing}

\author[a]{Arhum Ansari,}
\author[a]{Deep Mazumdar}
\author[a]{and Brijesh Thakkar}

\affiliation[a]{Indian Institute of Science Education and Research,\\ Dr Homi Bhabha Road, Pashan, Pune, India}

\emailAdd{ansari.arhum@students.iiserpune.ac.in,\\deepkamal.mazumdar@students.iiserpune.ac.in,\\brijesh.thakkar@students.iiserpune.ac.in}

\abstract{Using the recently developed framework in which cosmological correlators in (E)AdS are represented as flat-space amplitudes dressed by auxiliary propagators, we show that several structural properties of the cosmological observables have a direct flat-space origin. We derive cosmological cutting rules for spinning correlators from the flat-space optical theorem, obtain the cosmological tree theorem from the Feynman tree theorem, and uplift BCFW recursion relations to (E)AdS via dressing. We also show that flat-space soft theorems reproduce the soft limits of (E)AdS correlators, and find indications of an emergent universal structure in subleading soft limits. These results provide evidence that key features of cosmological correlators can be systematically understood as dressed manifestations of flat-space physics.}

\begin{document}
\maketitle

\section{Introduction}\label{sec:intro}

De Sitter (dS) space plays a central role in modern theoretical physics due to both observational and conceptual motivations. Cosmological observations suggest that the early universe underwent an inflationary phase, which can be well approximated by dS geometry. Moreover, our universe may again approach a dS-like phase at late times, making dS space the natural setting for understanding primordial fluctuations. That being said, dS space poses significant theoretical challenges: unlike flat space or AdS, it lacks global time translation symmetry and a timelike boundary, complicating the definition of observables and holography. As a result, observables are described in terms of late-time correlators or the wavefunction of the universe rather than scattering amplitudes, leading to considerably more intricate structures. These features make dS space an important arena for extending modern ideas such as holography and bootstrap methods to cosmological settings \cite{Arkani-Hamed:2018kmz,Baumann:2019oyu}.

Recent developments have revealed a surprising structural simplification of de Sitter correlators, showing that they can often be expressed in terms of flat-space scattering amplitudes, dressed with additional auxiliary propagators or time-dependent kernels. In this framework, the nested time integrals that typically arise in de Sitter calculations can be reorganized so that the essential dynamical information is carried by familiar flat-space amplitudes, while the effects of the expanding background are captured by theory-dependent auxiliary propagators \cite{Chowdhury:2025ohm, Chowdhury:2025nnk}. This representation provides a direct bridge between cosmological observables and flat-space physics, suggesting that many of the analytic properties and consistency conditions of amplitudes can be naturally extended to de Sitter \cite{Ansari:2026xkm, Chowdhury:2026upp}.

More recently \cite{Ansari:2026xkm}, it has been understood that a key structural property of flat-space amplitudes, the optical theorem, can be appropriately dressed and uplifted to cosmological observables in de Sitter space. In this framework, unitarity of time evolution gives rise to discontinuity relations for the wavefunction coefficients, which are precisely the cosmological cutting rules that have been independently discovered in the dS context \cite{Goodhew:2020hob, Melville:2021lst, Goodhew:2021oqg}, (see also \cite{Das:2025qsh,Colipi-Marchant:2025oin,Ansari:2026xkm,Chowdhury:2026upp,Das:2026vfv} for recent developments in this direction). These relations connect higher-point correlators to products of lower-point functions and provide a direct analogue of flat-space cutting rules, thereby offering powerful constraints on the analytic structure of cosmological observables despite the absence of a conventional S-matrix \cite{Marolf:2012kh, Melville:2023kgd, Melville:2024ove}.

These developments naturally raise the broader question of whether other structural features of de Sitter observables can also be understood as arising from flat-space physics dressed with appropriate auxiliary propagators \cite{Raju:2010by,Armstrong:2020woi,Albayrak:2020fyp}. In this work, we provide further evidence for this perspective. We show that the cosmological tree theorem \cite{AguiSalcedo:2023nds} can be obtained by dressing the Feynman tree theorem, and the AdS recursion relation \cite{Raju:2010by} are dressed avatar of flat-space BCFW recursion relation. Furthermore, we illustrate that the flat-space soft theorems for gauge theories uplift naturally to the soft limits of (EA)dS correlators. Moreover, while a universal structure for subleading soft limits in (EA)dS has remained elusive, we find indications that such a structure can emerge once appropriate dressing factors are incorporated.
 
The paper is organized as follows: In section \ref{sec:dress}, we review the notion of dressing for wavefunction coefficients with an example and provide the dressing rules for class of theories considered in this work. In section \ref{sec:Cutting}, starting with flat space cutting rules, we derive cutting rules for spinning theories in (EA)dS via dressing. In section \ref{sec:Tree}, we start with a brief review of Feynman Tree theorem in flat space and its cosmological counterpart in dS, and then establish the connection between them through dressing. We then uplift the flat-space notion of the BCFW recursion relation in section \ref{BCFW}, before finally uplifting the flat-space (sub-)leading soft theorems to their (EA)dS counterparts in section \ref{Soft}.

We supplement this paper with some essential appendices. We lay out the notations and conventions in appendix \ref{app:note}. We then briefly discuss the cutting rules for in-in correlators in appendix \ref{app:cutinin}. Thereafter, we present the dressing of different flat-space propagators in appendix \ref{app:dressprop}. Lastly, in appendix \ref{ap:calc} we present the calculation details of dressing the flat space four-point 1-loop amplitude for scalar theory.

\section{Dressing Rules for Theories of Interest}\label{sec:dress}
In this section, we briefly review the dressing procedure that uplifts flat-space Feynman diagrams to Witten diagrams in (EA)dS space \cite{Chowdhury:2025ohm}, which are related to the wavefunction coefficient in cosmology. We also specify the precise form of these theory-dependent dressing factors, which we will use frequently throughout the paper.

As an illustrative example, consider the six-point tree-level diagram in conformally coupled $\phi^4$ theory. The s-channel contribution to the wavefunction coefficient is 
\begin{align}
\psi_{6,s}(\{k_i\},\vec{s})
&=
\int dp
\Bigg(
\frac{p}{p^2+k_{123}^2}
\Bigg)
\Bigg(
\frac{p}{p^2+k_{456}^2}
\Bigg)
\Bigg(
\frac{g^2}{p^2+s^2}
\Bigg),
\label{eq:result6pt}
\end{align}
where $k_{abc}=k_a+k_b+k_c$ and $\vec{s}=\vec{k}_1+\vec{k}_2+\vec{k}_3$.

The final factor in \eqref{eq:result6pt} is  the flat-space amplitude for massless theory in Euclidean signature with relaxed energy-conservation, while the remaining factors arise from the bulk radial integrals and constitute the dressing factors. These theory-dependent dressings uplift flat-space amplitudes to wavefunction coefficients (and cosmological correlators) \cite{Chowdhury:2025ohm,Chowdhury:2025nnk}.

That being said, we will restrict ourselves to the computation of wavefunction coefficients, which are computed using Witten diagrams where fields obey Dirichlet boundary condition. We now enumerate the various dressing factors that will be used in later sections.
\begin{enumerate}
    \item Conformally coupled $\phi^4$
    \begin{equation}\label{eq:phi4dressing}
      \hat{\Delta}^{T,L}_{Y}(k_{ext},p)=\int dz\; e^{-k_{ext}z}\;\textrm{sin}(pz)=\frac{p}{p^2+k_{ext}^2},
    \end{equation}
    where, $Y$ is the place holder for $\phi^4$, Scalar QED and Yang-Mills\footnote{In four dimensions, the dressing rules are same for confomally coupled $\phi^4$ thoery, Scalar QED and Yang-Mills.}. The labels (T) and (L) denote the transverse and longitudinal components, respectively, in the cases of scalar QED and Yang–Mills theory. 
    \item  Conformally coupled $\phi^3$
    \begin{equation}\label{eq:phi3dressing}
      \hat{\Delta}_{\phi^4}(k_{ext},p)=\int \frac{dz}{z}\; e^{-k_{ext}z}\;\textrm{sin}(pz).
    \end{equation}
    \item Massless scalar minimally coupled to gravity
    \begin{equation}\label{eq:gravitydressing}
    \hat{\Delta}^{TT,L}_{GR}(k_{ext},p)=\int \frac{dz}{\sqrt{z}}\; \prod_{i=1}^{n} (1+k_i z)\, \,e^{-k_{\text{ext}}z}\;\sqrt{p}\, J_{\frac{3}{2}} (p z),
    \end{equation}
    where, $n$ is the number of external legs and $J_\nu$ is the Bessel function of first kind.  The labels (TT) and (L) denote the transverse-traceless and longitudinal components, respectively.
\end{enumerate}

Equipped with the necessary toolkit, we will now discuss on how to uplift various aspects of scattering amplitudes to (EA)dS correlators via the dressing procedure. We start with the cutting rules.


\section{Cutting Rules for Spinning fields}\label{sec:Cutting}
In this section, we show how the flat-space optical theorem can be uplifted to derive the cosmological cutting rules for various theories involving spinning fields. All results presented in this section are formulated in terms of wavefunction coefficients\footnote{An analogous analysis can be carried out for in-in correlators; the corresponding derivation is presented in the appendix\,\ref{app:cutinin}.}. This is a generalization of the results obtained for scalars in\ \cite{Ansari:2026xkm}. Let us quickly review the optical theorem in flat space before we move on to discuss cutting rules in (EA)dS via dressing.

\noindent Unitarity in the flat-space scattering amplitude implies,
\begin{align}\label{eq:ssdagger}
S S^{\dagger} = \mathbf{I} .
\end{align}
Writing \( S = \mathbf{I} + i T \), this translates to
\begin{align}\label{eq:ttdagger}
-\,i\,(T - T^{\dagger}) = T T^{\dagger} .
\end{align}
We can also write \eqref{eq:ttdagger} in terms of matrix elements as follows,
\begin{align}\label{eq:opticalthm}
-\,i\!\left[
\mathcal{A}(a \to b) - \mathcal{A}^{*}(b \to a)
\right]
=
\sum_{f}
d\Pi_f\,
\mathcal{A}^{*}(b \to f)\,
\mathcal{A}(a \to f) ,
\end{align}
where \(a\) and \(b\) are asymptotic states and \(f\) is all possible intermediate
states and \(d\Pi_f\) is the phase space measure.
The equation \eqref{eq:opticalthm} is  referred to as the \emph{optical theorem}.
For the case, when \(a\) and \(b\) are two-particle states, we get
\begin{align}\label{eq:opticalthm2p}
-\,i\!\left[
\mathcal{A}(k_1 k_2 \to p_1 p_2)
-
\mathcal{A}^{*}(p_1 p_2 \to k_1 k_2)
\right]
=
\sum_{n}
\left(
\prod_{i=1}^{n}
\int \frac{d^{3} q_i}{(2\pi)^3}\frac{1}{2|\vec{q}_i|} 
\right)
&\mathcal{A}^{*}(p_1 p_2 \to q_i)\,
\mathcal{A}(k_1 k_2 \to q_i)
\nonumber\\
&\quad\times
(2\pi)^{4}\,
\delta^{4}\!\left(k_1 + k_2 - \{q_i\}\right) ,
\end{align}
where $n$ runs over all possible intermediate particle states.

We will see that dressing the flat-space optical theorem with appropriate dressing rules leads to discontinuity relations among the correlators of corresponding theories in (EA)dS space. We first illustrate the dressing procedure through several tree-level examples.
\subsection{Tree level}

In this section, we examine a set of representative examples at tree level, where the essential
structures and physical principles can be seen in their simplest and most transparent form. We will discuss the following theories involving spinning fields to demonstrate the idea: Scalar QED, Yang Mills and scalar field  coupled to gravity.

\subsubsection*{Scalar QED}

One of the simplest theories involving spinning gauge fields is the scalar QED. Consider the s-channel four-point tree-level diagram, where the imaginary part of the amplitude is \footnote{Throughout this work, we will restrict ourselves to the `spatial three-dimensions' in tensor structures of the amplitude when uplifting amplitudes to boundary correlators in (EA)dS.}
\begin{align}\label{eq:IMscalarQEDamplitude}
2\,\mathrm{Im}\,\mathcal{A}_{4,s}^{\text{SQED}}
&=
2\pi\,g^{2}\,\alpha^i \beta^j \pi_{ij}\,\delta\!\left(p^{2}-s^{2}\right)\nonumber  \\
&= 2\pi\,g^{2}\,\alpha^i \beta^j \frac{\pi_{ij}}{2\abs{\vec{s}}}\,\left(\delta(p-\abs{\vec{s}})+\delta(p+\abs{\vec{s}})\right),
\end{align}
where we consider only the positive-energy delta function and retain the first term.\\
Here, $\pi_{ij} = \eta_{ij}- \frac{s_i s_j}{s^2}  $, $\alpha^i = (\vec{k}_1-\vec{k}_2)^i$,  $\beta^j = (\vec{k}_3-\vec{k}_4)^j$, $p = |\vec{k}_1|+|\vec{k}_2|$ is the total energy entering the vertex and $\vec{s} = \vec{k}_1+\vec{k}_2$ is total spatial momenta\footnote{\noindent We are working in covariant gauge ($\zeta = 0$) for the photon propagator. The photon propagator is   $\frac{1}{p^2+i \epsilon}\left(\eta_{ij}-\left(1-\zeta\right)\frac{p_i\,p_j}{p^2}\right)$.}.

According to the optical theorem, the imaginary part is determined by the right-hand side of \eqref{eq:opticalthm2p}. Explicitly, the right-hand side of the optical theorem in \eqref{eq:opticalthm2p} takes the following form,
\begin{equation}\label{eq:RHSopt}
    \sum_h \int \frac{d^3 \vec{q}}{(2\pi)^3}\,\frac{1}{2|\vec{q}|}\,\left(g\alpha^i \epsilon^h_i(\vec{q})\right)\, \left(g\beta^j \epsilon^{*\,h}_j(\vec{q})\right)\,
(2\pi)^4\,\delta^{4}(k_1 + k_2 - q)\,
=
\frac{g^2\alpha^i \beta^j \pi_{ij}}{2\abs{\vec{s}}}\,
(2\pi)\delta(p - |\vec{s}|),
\end{equation}
where we used the completeness relation for polarization vectors given as,
\begin{equation}\label{eq:completeness}
    \pi_{ij}=\sum_h \epsilon^h_i(\vec{s}) \epsilon^{{*}\,h}_j (\Vec{s}) = \eta_{ij} - \frac{s_i s_j}{s^2}.
\end{equation}
We will now uplift this amplitude to (EA)dS space using the dressing rules given in section \ref{sec:dress}. When lifting the flat-space amplitude to(EA)dS, we relax the energy conservation and  analytically continue
the \(s\) variable as \(s \to -i s\).
We will use this convention whenever we lift the flat-space amplitude
to get(EA)dS correlators.

After analytic continuation, we now dress both sides of \eqref{eq:opticalthm2p} with the
appropriate dressing factors to get the analog of optical theorem for boundary correlators (wavefunction coefficients) in (EA)dS. Applying dressing as in \eqref{eq:phi4dressing} on  transverse part\footnote{While the longitudinal part of the propagator does contribute to the correlator, its discontinuity is zero. Henceforth, we will not present the action of cut on the longitudinal part. Please refer to appendix \ref{app:cutinin} for more details on this.} of \eqref{eq:IMscalarQEDamplitude} we get
\begin{align}
   & g^{2}\,\alpha^i \beta^j \pi_{ij}\, \int_{-\infty}^{\infty}
dp\,\left(\int_{0}^{\infty} dz_1\,dz_2\,
e^{-k_{12} z_1 - k_{34} z_2}\,
\mathrm{sin}(p z_1)\,\mathrm{sin}(p z_2)\right)2\pi\,\delta\!\left(p^{2}+s^{2}\right)  \nonumber \\
& = -i \, \mathrm{Disc}_s \left[\psi^T_{4,s}(\left\{k_i\right\},\vec{s})\right].
\end{align}
Performing same dressing on \eqref{eq:RHSopt}, we obtain
\begin{align}
    &g^2\alpha^i \beta^j \pi_{ij}\, \int_{-\infty}^{\infty} \frac{dp}{2\abs{\vec{s}}}\, \left(\int dz_1\, e^{-k_{12} z_1}\,\mathrm{sin}(p z_1)\right)\,\left(\int dz_2\, e^{-k_{34} z_2}\,\mathrm{sin}(p z_2)\right)
2\pi\,\delta(p+is)\,\nonumber \\
&= - \sum_h  \frac{1}{2\abs{\vec{s}}}\left( g\,\alpha^{i}\epsilon^h_i(\vec{s}) 
\int_{0}^{\infty} dz_1\, e^{-k_{12} z_1}\,\mathrm{sinh}(s\, z_1)\right)\,\left(g\, \beta^{j}  \epsilon^{*\,h}_j(\Vec{s}) \,\int_{0}^{\infty} dz_2\, e^{-k_{34} z_2}\,\mathrm{sinh}(s\,z_2)\right)\nonumber \\
& =  \sum_h \frac{1}{2\abs{\vec{s}}} \left(\mathrm{disc}_s  \left[\psi^h_{3}(\left\{k_L\right\},\vec{s})\right]\right)\,\left({\mathrm{disc}}_s \left[\psi^h_{3}(-\vec{s},\left\{k_R\right\})\right]\right),
\end{align}
where, $K_L$ and $K_R$ denote external energies for left and right vertex. We have used the hermitian analyticity property of polarization vectors, i.e. $\epsilon_i(-s) = \epsilon_i^{*}(s)$. Hence, we see that the flat-space optical theorem, as given in \eqref{eq:opticalthm2p}, implies the following ,
\begin{equation}\label{eq:discrela}
    \mathrm{Disc}_s \left[\psi^T_{4,s}(\left\{k_i\right\},\vec{s})\right] =  i \sum_h \frac{1}{2\abs{\vec{s}}} \left(\mathrm{disc}_s  \left[\psi^h_{3}(\left\{k_L\right\},\vec{s})\right]\right)\,\left({\mathrm{disc}}_s \left[\psi^h_{3}(-\vec{s},\left\{k_R\right\})\right]\right).
\end{equation}
We have used the definition of ``Disc" and ``disc" as defined in \eqref{eq:Disc} and \eqref{eq:disc}.
Note that this is in agreement with the result obtained in \cite{Goodhew:2021oqg}. Hence, we see that uplifting the flat-space optical theorem directly reproduces the discontinuity relations in (EA)dS space.

\subsubsection*{Yang Mills theory}

We now turn to the case of Yang–Mills theory, where we extend the preceding analysis to non-Abelian gauge interactions and examine how the corresponding structures manifest in this more intricate setting. Considering the s-channel color-ordered four-point amplitude at tree level, the imaginary part of amplitude is written as,
\begin{equation}\label{eq:IMYMamplitude}
2\,\mathrm{Im}\,\mathcal{A}_{4,s}^{\text{JJJJ}}
=
2\pi\,g^2\,V^{i}_{12} V^{j}_{34}\, \pi_{ij}\,\delta\!\left(p^{2}-s^{2}\right)\;,
\end{equation}
where,
\begin{equation}
V^{\,i}_{ab}
\equiv
\epsilon_a \!\cdot\! \epsilon_b \,(\vec k_a-\vec k_b)^{i}
+2(\epsilon_a\!\cdot\!\vec k_b)\,\epsilon_b^{\,i}
-2(\epsilon_b\!\cdot\!\vec k_a)\,\epsilon_a^{\,i}.
\end{equation}
Here $a,b$ are particle indices and $\pi_{ij}(s)= \eta_{ij} - \frac{s_i s_j}{s^2}$. Which makes up the LHS of \eqref{eq:opticalthm2p}. The RHS of the optical theorem as given in \eqref{eq:opticalthm2p} is given as,
\begin{equation}\label{eq:RHSoptYM}
    \sum_h \int \frac{d^3 \vec{q}}{(2\pi)^3}\,\frac{1}{2|\vec{q}|}\,\left(gV^{i}_{12}\epsilon^h_i(\vec{q})\right)\, \left(gV^{j}_{34} \epsilon^{*\,h}_j(\vec{q})\right)\,
(2\pi)^4\,\delta^{4}(k_1 + k_2 - q)\,
=
\frac{g^2\,V^{i}_{12} V^{j}_{34}\, \pi_{ij}}{2\abs{\vec{s}}}\,
(2\pi)\delta(p - |\vec{s}|)\,,
\end{equation}
where we have used,
\begin{equation}
      \mathcal{A}_3^{\textrm{YM}} = g V^i_{ab}\,\epsilon_i .
\end{equation}
Now we uplift \eqref{eq:opticalthm2p} to (EA)dS space by applying appropriate dressing. 
Applying dressing as in \eqref{eq:phi4dressing} on  transverse part of \eqref{eq:IMYMamplitude} we get,
\begin{align}
   & g^2\,V^{i}_{12} V^{j}_{34}\,\pi_{ij}\, \int_{-\infty}^{\infty}
dp\,\left(\int_{0}^{\infty} dz_1\,dz_2\,
e^{-k_{12} z_1 - k_{34} z_2}\,
\mathrm{sin}(p z_1)\,\mathrm{sin}(p z_2)\right)(2\pi)\delta\!\left(p^{2}+s^{2}\right)  \nonumber \\
& = -i\,\mathrm{Disc}_s \left[\psi^{JJJJ}_{4,s}(\left\{k_i\right\},\vec{s})\right].
\end{align}
Performing same dressing on \eqref{eq:RHSoptYM}, we obtain
\begin{align}
    &g^2\,V^{i}_{12} V^{j}_{34}\, \pi_{ij}\, \int_{-\infty}^{\infty} \frac{dp}{2\abs{\vec{s}}}\, \left(\int dz_1\, e^{-k_{12} z_1}\,\mathrm{sin}(p z_1)\right)\,\left(\int dz_2\, e^{-k_{34} z_2}\,\mathrm{sin}(p z_2)\right)
\delta(p + is)\,\nonumber \\
&=-  \sum_h  \frac{1}{2\abs{\vec{s}}}\left( g\,V_{12}^{\;i}\epsilon^h_i(\vec{s}) 
\int_{0}^{\infty} dz_1\, e^{-k_{12} z_1}\mathrm{sinh}(s\,z_1)\right)\,\left(g\, V_{34}^{\;j}  \epsilon^{*\,h}_j(\Vec{s}) \,\int_{0}^{\infty} dz_2\, e^{-k_{34} z_2}\mathrm{sinh}(s\,z_2)\right)\nonumber \\
& =  \sum_h \frac{1}{2\abs{\vec{s}}} \left({\mathrm{disc}}_s \left[\psi^{JJJ}_3(\left\{k_L\right\},\vec{s})\right]\right)\,\left({\mathrm{disc}}_s  \left[\psi^{JJJ}_3(-\vec{s},\left\{k_R\right\})\right]\right).
\end{align}
This, in turn, implies that the flat-space optical theorem leads to, once again, the same discontinuity relation \eqref{eq:discrela} we obtained earlier.

Having discussed Yang–Mills theory, we now turn to an example involving the graviton field, where we extend the preceding analysis to gravitational interactions and examine how the corresponding structures manifest in this setting.

\tikzfeynmanset{
  graviton/.style={
    double,
    decorate,
    decoration={snake, amplitude=2pt, segment length=6pt}
  }
}

\subsubsection*{Scalar with graviton exchange}

Consider a massless scalar field minimally coupled to gravity. The tree-level s-channel four-point amplitude for scalars mediated by graviton exchange. The imaginary part of amplitude is written as,
\begin{equation}\label{eq:IMGRamplitude}
2\,\mathrm{Im}\,\mathcal{A}_4^{\text{Tree}}
=
2\pi\,g^2\,V^{ij}_{12} V^{lm}_{34}\, \Pi_{ijlm}\,\delta\!\left(p^{2}-s^{2}\right),
\end{equation}
where, 
\begin{align}
V_{ab}^{ij}&=k_{a,i}\,k_{b,j}+k_{a,j}\,k_{b,i},\notag\\
\Pi_{ijkl}=\frac{1}{2}\sum_{h}\epsilon^{\,h}_{ij}(\vec{k})\,\epsilon^{\,h}_{kl}(-\vec{k})&
=\frac{1}{2}\,\pi^{k}_{ik}\,\pi^{k}_{jl}+\frac{1}{2}\,\pi^{k}_{il}\,\pi^{k}_{jk}-\frac{1}{2}\,\pi^{k}_{ij}\,\pi^{k}_{kl}.
\end{align}
which constitutes the LHS of \eqref{eq:opticalthm2p} while the RHS is given as,
\begin{equation}\label{eq:RHSoptgraviton}
    \sum_h \int \frac{d^3 \vec{q}}{(2\pi)^3}\,\frac{1}{2|\vec{q}|}\,\left(g V^{ij}_{12}\epsilon^h_{ij}(\vec{q})\right)\, \left(g V^{lm}_{34} \epsilon^{*\,h}_{lm}(\vec{q})\right)\,
(2\pi)^4\,\delta^{4}(k_1 + k_2 - q)\,
=
\frac{g^2\,V^{ij}_{12} V^{lm}_{34}\, \Pi_{ijlm}}{2\abs{\vec{s}}}\,
(2\pi)\delta(p - |\vec{s}|).
\end{equation}
Uplifting Eq.\,\eqref{eq:opticalthm2p} to (EA)dS leads to the corresponding discontinuity relations for boundary correlators (or wavefunction coefficients). Applying dressing as in \eqref{eq:gravitydressing} on  transverse part of \eqref{eq:IMGRamplitude} we get,
\begin{align}
 & g^2\,V^{ij}_{12} V^{lm}_{34}\,\Pi_{ijlm}
\int_{-\infty}^{\infty} dp\;
\Bigg(
\int_{0}^{\infty} dz_1\,dz_2\;
e^{-k_{12} z_1 - k_{34} z_2}\,(z_1 z_2)^{-1/2}
\left(1+k_1 z_1\right)\left(1+k_2 z_1\right)
\nonumber\\
& \qquad \qquad \qquad \qquad \qquad
\times\,\left(1+k_3 z_2\right)\left(1+k_4 z_2\right)\, p\, J_{3/2}(p z_1)\,J_{3/2}(p z_2)
\Bigg)\,
(2\pi)\delta\!\left(p^{2}+s^{2}\right) \nonumber \\
&= -i \, \mathrm{Disc}_s \left[\psi^{TT}_{4,s}(\left\{k_i\right\},\vec{s})\right].
\end{align}
Performing same dressing on \eqref{eq:RHSoptgraviton}, we obtain
\begin{align}
    &g^2\,V^{ij}_{12} V^{lm}_{34}\, \Pi_{ijlm}\, \int_{-\infty}^{\infty} \frac{dp}{2\abs{\vec{s}}}\, \left(\int dz_1\, e^{-k_{12} z_1}\,\sqrt{p}\,z_1^{-1/2}\,\left(1+k_1 z_1\right)\left(1+k_2 z_1\right)\,J_{3/2}(p z_1)\,\right)\nonumber \\
     &\qquad \qquad \qquad \qquad \qquad  \times \left(\int dz_2\, e^{-k_{34} z_2}\,\sqrt{p}\,z_2^{-1/2}\,\left(1+k_3 z_2\right)\left(1+k_4 z_2\right)\,J_{3/2}(p z_2)\,\right)\,(2\pi)\delta(p + is)\,\nonumber \\
& =   \sum_h \frac{1}{2\abs{\vec{s}}} \left(\mathrm{disc}_s  \left[\psi^h_{3}(\left\{k_L\right\},\vec{s})\right]\right)\,\left({\mathrm{disc}}_s \left[\psi^h_{3}(-\vec{s},\left\{k_R\right\})\right]\right).
\end{align}
and hence,
\begin{equation}
    \mathrm{Disc}_s \left[\psi^{TT}_{4,s}(\left\{k_i\right\},\vec{s})\right] =   i \sum_h \frac{1}{2\abs{\vec{s}}} \left(\mathrm{disc}_s  \left[\psi^h_{3}(\left\{k_L\right\},\vec{s})\right]\right)\,\left({\mathrm{disc}}_s \left[\psi^h_{3}(-\vec{s},\left\{k_R\right\})\right]\right).
\end{equation}
This agrees with the result obtained independently in the context of (EA)dS \cite{Goodhew:2021oqg}.

We now turn to loop-level processes and examine how the flat-space optical theorem, upon appropriate uplift, gives rise to discontinuity relations for loop correlators in (EA)dS.


\subsection{Loops}

In this section, we show how the flat-space optical theorem at loop level can be uplifted to derive discontinuity relations in (EA)dS. We illustrate this through several examples, beginning with scalar QED.

\subsubsection*{Scalar QED}

\usetikzlibrary{decorations.pathmorphing}

\begin{figure}[h]
\centering
\begin{tikzpicture}[thick]

\node at (-2,1.4) {$\vec{k}_1$};
\node at (-2,-1.4) {$\vec{k}_2$};

\node at (2,1.4) {$\vec{k}_3$};
\node at (2,-1.4) {$\vec{k}_4$};

\coordinate (v1) at (-1,0);
\coordinate (v2) at (1,0);

\draw (-2,1) -- (v1);
\draw (-2,-1) -- (v1);

\draw (2,-1) -- (v2);
\draw (2,1) -- (v2);

\draw[line width = 0.7pt,decorate, decoration={snake,amplitude=1pt, segment length=5pt}, bend left=30] (v1) to (v2);
\draw[line width = 0.7pt,decorate, decoration={snake, amplitude=1pt, segment length=5pt}, bend right=30] (v1) to (v2);

\node at (0,0.8) {$\vec{l}_1$};
\node at (0,-0.8) {$\vec{l}_2$};

\end{tikzpicture}
\caption{four-point s-channel diagram with photon bubble loop.}\label{fig:SQED4p1l}
\end{figure}
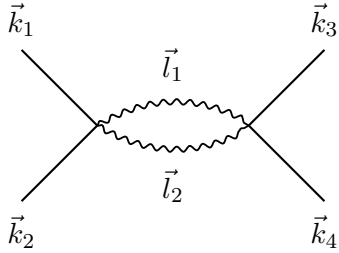

Consider the four-point bubble diagram given in\,Fig.\,\ref{fig:SQED4p1l}, with external legs scalar and photon running in the loop. At loop level, the optical theorem implies the following relation between the imaginary part of the amplitude and on-shell intermediate contributions,
\begin{align}\label{eq:imA4loop}
&2\mathrm{Im}(\mathcal{A}_{4,s}^{1-loop})\notag\\
&= \int\, \frac{d^3 l_1}{2|\vec{l}_1|}\,\frac{d^3 l_2}{2|\vec{l}_2|}\, \mathcal{A}_4(k_1k_2\rightarrow l_1l_2)\,\mathcal{A}^*_4(k_3k_4\rightarrow l_1l_2)\,(2\pi)^4\delta^4(K_1+K_2+L_1-L_2)\,\nonumber \\
 &= \sum_{h,h'}\int\, \frac{d^3 l_1}{2|\vec{l}_1|}\,\frac{d^3 l_2}{2|\vec{l}_2|}\, \left(g^2 \delta^{ij} \,\epsilon^h_i({\vec{l}_1})\,\epsilon^{h'}_j({\vec{l}_2})\,\right) \left(g^2\delta^{lm} \,\epsilon_l^{*h}({\vec{l}_1})\,\epsilon_m^{*h'}({\vec{l}_2})\,\right)\,(2\pi)^4\delta^4(K_1+K_2+L_1-L_2)\,\nonumber \\
     & = g^4\, \int\, d^3 l_1\,d^3 l_2\,\frac{dp_1}{2p_1}\,\frac{dp_2}{2p_2} \pi_{jl}(\vec{l}_1)\,\pi_{jl}(\vec{l}_2)\,(2\pi)^2\delta(p_1-|\vec{l}_1|\,\delta(p_2-|\vec{l}_2|)\,(2\pi)^4\delta^4(K_1+K_2+L_1-L_2),
\end{align}
where, $K_i = (k,\vec{k}_i)$, and $L_i = (p,\vec{l}_i)$. 

Imaginary part of the amplitude for this process is given as,
\begin{align}\label{eq:lhsoptSQED4p1l}
    &2\mathrm{Im}(\mathcal{A}_4^{1-loop})\notag\\
    &= g^4\,\int d^3 l_1\,d^3 l_2\, dp_1\, dp_2 \, \pi_{jl}(\vec{l}_1)\,\pi_{jl}(\vec{l}_2)\,(2\pi)^2\delta(p_1^2- l_1^2)\,\delta(p_2^2-l_2^2)\,(2\pi)^4\delta^4(K_1+K_2+L_1-L_2).
\end{align}
Using the following decomposition
\begin{equation}
    \delta(p^2-l^2) = \frac{1}{2|\vec{l}|}\left( \delta(p-|\vec{l}|)+ \delta(p+|\vec{l}|)\right),
\end{equation}
we  get the expression same as \eqref{eq:imA4loop} (where we have excluded negative energies). Moreover, we will work only with the
integrands for loops, which is defined as follows,
\begin{equation}
    \psi_4(\left\{k_i\right\}) = \int d^3l_1\, d^3l_2\, \tilde{\psi}_4(\left\{k_i\right\},\vec{l}_1,\vec{l}_2).
\end{equation}

We now proceed to obtain the corresponding (EA)dS correlator by dressing the flat-space amplitude with the appropriate dressing factors.

Dressing \eqref{eq:lhsoptSQED4p1l} with dressing factors as in \eqref{eq:phi4dressing} we get, 
\begin{align}\label{LHSloop1}
   & \pi_{jl}\,\pi_{jl}\, \int\, dp_1\, dp_2\, dz_1\, dz_2\, e^{-k_{12} z_1}\, \mathrm{sin}(p_1 z_1)\,  \mathrm{sin}(p_1 z_2)\,e^{-k_{34} z_2}\, \mathrm{sin}(p_2 z_1)\,  \mathrm{sin}(p_2 z_2)\,(2\pi)^2  \delta(p_1^2+ l_1^2)\,\delta(p_2^2+l_2^2)\nonumber \\
   & = - \mathrm{Disc}_{l_1} \mathrm{Disc}_{l_2} \left[\tilde{\psi}^{\mathrm{1-loop}}_{4,s}(\left\{k_i\right\},\vec{l}_1,\vec{l}_2)\right].
\end{align}
Now, dressing \eqref{eq:imA4loop},  we get,
\begin{align}
     & \pi_{jl}\,\pi_{jl}\, \int\, \frac{dp_1}{2p_1}\,\frac{dp_2}{2p_2}\,\int dz_1\, e^{-k_{12} z_1}\, \mathrm{sin}(p_1 z_1)\,  \mathrm{sin}(p_2 z_1)\,\nonumber \\
     &\qquad \qquad \qquad \qquad  \times \int dz_2\,e^{-k_{34} z_2}\, \mathrm{sin}(p_1 z_2)\,  \mathrm{sin}(p_2 z_2)\, (2\pi)^2 \delta(p_1 + i l_1)\,\delta(p_2+i l_2)\nonumber \\
     & = -\sum_{h,h'}\frac{1}{4|\vec{l}_1||\vec{l}_2|} \int\, dz_1\,\left(\delta^{ij}\epsilon^h_i(\vec{l}_1)\epsilon^{h'}_j(\vec{l}_2)e^{-k_{12} z_1}\, \mathrm{sinh}(l_1 z_1)\,  \mathrm{sinh}(l_2 z_1)\right)\,\nonumber \\
     & \qquad \qquad \qquad \qquad \qquad \qquad \times \int dz_2\,\left(\delta^{lm}\epsilon^{*\,h}_l(\vec{l}_1)\epsilon^{*\,h'}_m(\vec{l}_2)e^{-k_{34} z_2}\, \mathrm{sinh}(l_1 z_2)\,  \mathrm{sinh}(l_2 z_2)\right)\,\nonumber\\
     & =  -\sum_{h,h'}\frac{1}{4|\vec{l}_1||\vec{l}_2|} \left(\mathrm{disc}_{l_1} \mathrm{disc}_{l_2} \left[\tilde{\psi}_{4}(\left\{k_L\right\},\vec{l}_1,\vec{l}_2)\right]\right)  \left(\mathrm{disc}_{l_1} \mathrm{disc}_{l_2} \left[\tilde{\psi}_{4}(-\vec{l}_1,-\vec{l}_2,\left\{k_R\right\})\right]\right) .
\end{align}\label{RHSloop1}
Comparing \eqref{LHSloop1} and \eqref{RHSloop1} leads to the following relation
\begin{align}\label{eq:loopsresult}
&\mathrm{Disc}_{l_1} \mathrm{Disc}_{l_2} \left[\psi^{\mathrm{1-loop}}_{4,s}(\left\{k_i\right\},\vec{l}_1,\vec{l}_2)\right]
\notag\\
&= \sum_{h,h'}\frac{1}{4|\vec{l}_1||\vec{l}_2|} \left(\mathrm{disc}_{l_1} \mathrm{disc}_{l_2} \left[\psi_{4}(\left\{k_L\right\},\vec{l}_1,\vec{l}_2)\right]\right)
\times\left(\mathrm{disc}_{l_1} \mathrm{disc}_{l_2} \left[\psi_{4}(-\vec{l}_1,-\vec{l}_2,\left\{k_R\right\})\right]\right),
\end{align}
which is in agreement with the results obtained for cosmological cutting rules \cite{Melville:2021lst}. We now turn to Yang–Mills theory and examine how the optical theorem at loop level gives rise to cutting rules for correlators in (EA)dS.


\subsubsection*{Yang-Mills theory}

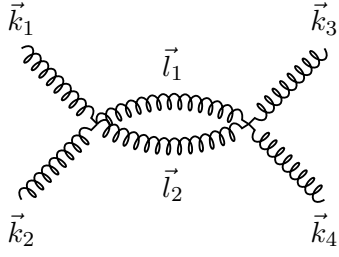
\begin{figure}[h]
\centering
\begin{tikzpicture}[thick]

\node at (-2,1.4) {$\vec{k}_1$};
\node at (-2,-1.4) {$\vec{k}_2$};

\node at (2,1.4) {$\vec{k}_3$};
\node at (2,-1.4) {$\vec{k}_4$};

\coordinate (v1) at (-1,0);
\coordinate (v2) at (1,0);

\draw[line width = 0.7pt,decorate, decoration={coil,amplitude=3pt, segment length=4pt}] (-2,1) -- (v1);
\draw[line width = 0.7pt,decorate, decoration={coil,amplitude=3pt, segment length=4pt}] (-2,-1) -- (v1);

\draw[line width = 0.7pt,decorate, decoration={coil,amplitude=3pt, segment length=4pt}] (2,-1) -- (v2);
\draw[line width = 0.7pt,decorate, decoration={coil,amplitude=3pt, segment length=4pt}] (2,1) -- (v2);

\draw[line width = 0.7pt,decorate, decoration={coil,amplitude=3pt, segment length=4pt}, bend left=30] (v1) to (v2);
\draw[line width = 0.7pt,decorate, decoration={coil, amplitude=3pt, segment length=4pt}, bend right=30] (v1) to (v2);

\node at (0,0.8) {$\vec{l}_1$};
\node at (0,-0.8) {$\vec{l}_2$};

\end{tikzpicture}
\caption{four-point s-channel diagram with photon bubble loop.}\label{fig:YM4p1l}
\end{figure}

In this section, we discuss the cutting rules for the more intricate case of Yang–Mills theory.

Consider four-point gluon bubble diagram with gluon running in loop as given in Fig.\,\ref{fig:YM4p1l}. From optical theorem we can write,

\begin{align}
\label{eq:imaym}
2\mathrm{Im}(\mathcal{A}_{4,s}^{1-loop}) &= \int\, \frac{d^3 l_1}{2|\vec{l}_1|}\,\frac{d^3 l_2}{2|\vec{l}_2|}\, \abs{\mathcal{A}_3}^2\,(2\pi)^4\delta(k_1+k_2+p_1-p_2)\, \delta^3(\vec{k}_1+\vec{k}_2 +\vec{l}_1-\vec{l}_2)\nonumber \\
     &= \sum_{h,h'}\int\, \frac{d^3 l_1}{2|\vec{l}_1|}\,\frac{d^3 l_2}{2|\vec{l}_2|}\, \left( g^2\,V_{jklm}\,\epsilon_j({\vec{k}_1})\,\epsilon_k({\vec{k}_2})\,\epsilon^h_l({\vec{l}_1})\,\epsilon^{h'}_m({\vec{l}_2})\,\right)\nonumber \\ 
     &\qquad  \qquad\,\times\left(g^2\,V_{pqrs}\,{\epsilon^h_p}^{*}({\vec{l}_1})\,{\epsilon^{h'}_q}^{*}({\vec{l}_2})\,\epsilon_r({\vec{k}_3})\,\epsilon_s({\vec{k}_4})\,\right)\,(2\pi)^4\delta^4(K_1+K_2+L_1-L_2)\nonumber \\
      &= g^4\, \int\, d^3 l_1\,d^3 l_2\,\frac{dp_1}{2p_1}\,\frac{dp_2}{2p_2} \left(V_{jklm}\,\epsilon_j({\vec{k}_1})\,\epsilon_k({\vec{k}_2})\,\right) \pi_{lp}\,\pi_{mq}\left(V_{pqrs}\,\epsilon_r({\vec{k}_3})\,\epsilon_s({\vec{k}_4})\,\right)\,\nonumber \\& \qquad \qquad \qquad \qquad  \qquad \times (2\pi)^2\delta(p_1-l_1)\,\delta(p_2-l_2)\,(2\pi)^4\delta^4(K_1+K_2+L_1-L_2),
\end{align}
where,
\begin{equation}
    V_{jklm} = 2 \eta_{jl}\eta_{km}-\left( \eta_{jk}\eta_{lm}+ \eta_{jm}\eta_{kl}\right).
\end{equation}
The imaginary part of the one-loop amplitude takes the following form,
\begin{align}\label{eq:YmA41loopim}
   2\mathrm{Im}(\mathcal{A}_4^{1-loop}) = g^4\,\int d^3 l_1\,d^3 l_2\, dp_1\, dp_2 \, \mathcal{T}\,
    (2\pi)^2\delta(p_1^2- l_1^2)\,\delta(p_2^2-l_2^2)\,(2\pi)^4\delta^4(K_1+K_2+L_1-L_2),
\end{align}
We use the following short hand to denote tensor structure in above equation,
\begin{equation}
    \mathcal{T} = \epsilon_j(\vec{k}_1)\,\epsilon_k(\vec{k}_2)\,V_{jklm}\,\pi_{lp}\,\pi_{mq}\,V_{pqrs}\,\epsilon_{r}(\vec{k}_3)\,\epsilon_{s}(\vec{k}_4).
\end{equation}
Now to get the (EA)dS analogue, we dress \eqref{eq:YmA41loopim} and \eqref{eq:imaym}. Dressing the \eqref{eq:YmA41loopim} with dressing factors as in \eqref{eq:phi4dressing} gives,
\begin{align}\label{eq:part1YM}
   & \mathcal{T}\int\, dp_1\, dp_2\, dz_1\, dz_2\, e^{-k_{12} z_1}\, \mathrm{sin}(p_1 z_1)\,  \mathrm{sin}(p_1 z_2)\,e^{-k_{34} z_2}\, \mathrm{sin}(p_2 z_1)\,  \mathrm{sin}(p_2 z_2)\,(2\pi)^2  \delta(p_1^2+ l_1^2)\,\delta(p_2^2+l_2^2)\nonumber \\
   & = - \mathrm{Disc}_{l_1} \mathrm{Disc}_{l_2} \left[\tilde{\psi}^{\mathrm{1-loop}}_{4,s}(\left\{k_i\right\},\vec{l}_1,\vec{l}_2)\right].
\end{align}
Further, dressing \eqref{eq:imaym}, we get,
\begin{align}\label{eq:part2YM}
     & \mathcal{T} \int\, \frac{dp_1}{2p_1}\,\frac{dp_2}{2p_2}\,\int dz_1\, e^{-k_{12} z_1}\, \mathrm{sin}(p_1 z_1)\,  \mathrm{sin}(p_2 z_1)\,\int dz_2\,e^{-k_{34} z_2}\, \mathrm{sin}(p_1 z_2)\,  \mathrm{sin}(p_2 z_2)\, (2\pi)^2 \delta(p_1 + i l_1)\,\delta(p_2+i l_2)\nonumber \\
     & = -\mathcal{T}\, \int\, \frac{dz_1}{2|\vec{l}_1|}\,\frac{dz_2}{2|\vec{l}_2|}\, e^{-k_{12} z_1}\, \mathrm{sinh}(l_1 z_1)\,  \mathrm{sinh}(l_1 z_2)\,e^{-k_{34} z_2}\, \mathrm{sinh}(l_2 z_1)\,  \mathrm{sinh}(l_2 z_2)\,\nonumber\\
     & = -\sum_{h,h'}\frac{1}{4|\vec{l}_1||\vec{l}_2|} \left(\mathrm{disc}_{l_1} \mathrm{disc}_{l_2} \left[\tilde{\psi}_{4}(\left\{k_L\right\},\vec{l}_1,\vec{l}_2)\right]\right)  \left(\mathrm{disc}_{l_1} \mathrm{disc}_{l_2} \left[\tilde{\psi}_{4}(-\vec{l}_1,-\vec{l}_2,\left\{k_R\right\})\right]\right) .
\end{align}
Combining \eqref{eq:part1YM} and \eqref{eq:part2YM} we get the same result \eqref{eq:loopsresult} as obtained earlier.

The cutting rules express discontinuities of (EA)dS observables in terms of lower-point on-shell data, making the role of unitarity manifest at the level of correlators. Next, we discuss uplifting the flat-space Feynman tree theorem using appropriate dressing. Upon uplifting, this procedure reproduces the cosmological tree theorem discovered in the context of (EA)dS independently\ \cite{AguiSalcedo:2023nds}.


\section{Cosmological Tree Theorem from Feynman Tree theorem}\label{sec:Tree}

The Cosmological Tree Theorem (CTT) relates loop-level observables in dS/(EA)dS to sums over tree-level contributions, extending the logic of unitarity and cutting rules to cosmological settings\,\cite{AguiSalcedo:2023nds}. Analogous to the flat-space Feynman Tree Theorem (FTT), it expresses loop correlators or wavefunction coefficients in terms of on-shell data by systematically cutting internal propagators. In this section, we show that the CTT can be derived by appropriately dressing the FTT, thereby demonstrating how the principles of unitarity and causality in flat space uplift naturally to (EA)dS through suitable dressing factors.

We begin by reviewing the tree theorem in both flat space and de Sitter space, before outlining the dressing procedure and demonstrating how the cosmological tree theorem can be derived from its flat-space counterpart.

\subsection{Feynman Tree Theorem (FTT) in Flat Space}
The Feynman Tree Theorem  provides a systematic way to cut open any closed loop in a Feynman diagram and replace
it by a sum over simpler diagrams with fewer loops.
It comes from the following decomposition of the Feynman propagator,
\begin{align}\label{eq:GRdecompoflat}
\Pi_F
=
\frac{i}{p^2-s^2+i\epsilon}
=
\Pi_R
+
\frac{\pi}{s}\delta(p+s),
\end{align}
where, $p$ is the energy and $\vec{s}$ is the total momenta flowing into the propagator. The retarded propagator $\Pi_R$ is defined as
\begin{equation}\label{eq:GRflat}
    \Pi_{\text{R}} = \frac{i}{2s} \left[\frac{1}{p-s+i \epsilon}- \frac{1}{p+s+i \epsilon} \right],
\end{equation}

One can use the decomposition given in \eqref{eq:GRdecompoflat} to replace all the Feynman propagator in any given loop diagram. The presence of the $\delta$-function enforces on-shell conditions for internal lines, thereby allowing any loop diagram to be expressed as a sum over tree-level processes.

As an example, consider the one-loop four-point amplitude in $\phi^4$ theory,
\begin{equation}
i\mathcal{A}_{4,s}^{1\text{-loop}}
=
\frac{(i\lambda)^2}{2}
\int
\frac{d^4k}{(2\pi)^4}
\,
\Pi_F(k-p)\Pi_F(k).
\end{equation}
Using \eqref{eq:GRdecompoflat}, the terms containing only retarded propagators vanish by the choice of contour, and the amplitude reduces to
\begin{align}
i\mathcal{A}_{4,s}^{1\text{-loop}}
=
-\frac{\lambda^2}{2}
\int
\frac{d^4k}{(2\pi)^4}
\Bigg[
\Pi_F(k-p)\,
\frac{\pi}{\omega_k}\delta(k_0+\omega_k)
+
\Pi_F(k)\,
\frac{\pi}{\omega_{k-p}}
\delta(k_0-p_0+\omega_{k-p})
\Bigg],
\end{align}
where $\omega_k=\sqrt{|\vec{k}|^2+m^2}$. The $\delta$-functions localize the energy integral and place the internal propagators on shell, making explicit the decomposition of the loop amplitude into tree-level contributions.

\subsection{Cosmological Tree theorem (CTT) in dS space}\label{sec:CTTds}

The Cosmological tree theorem is analogous to the Feynman tree theorem, but for wavefunction coefficients instead of amplitudes. It can be used to express a Witten diagram with any number of loops in terms of purely tree-level diagrams \cite{AguiSalcedo:2023nds}. As an example, we consider the two-point one-loop diagram  for massless $\phi^3$ scalar theory in de-sitter space. The wavefunction coefficient for this is written as,
\begin{equation}\label{eq:2p1l}
    \psi^{1\text{-loop}}_{k_1 k_2}
    =
    \int_{q_1 q_2}
    K_{k_1}(\eta_1)\,
    G_{q_1}(\eta_1,\eta_2)\,
    G_{q_2}(\eta_1,\eta_2)\,
    K_{k_2}(\eta_2)\,
    \delta^3(\vec{k}_1+\vec{q}_1-\vec{q}_2),
\end{equation}
where $K_k$ and $G_k$ are the bulk-to-boundary and Feynman bulk-to-bulk propagators respectively.

Using CTT the loop wavefunction can be written entirely in terms of tree-level coefficients,
\begin{align}\label{eq:2p1lCTT}
\psi_{k_1 k_2}^{1\text{-loop}}
&=
\int_{q_1 q_1'} P_{q_1 q_1'}
\,\mathrm{disc}_{q_1'}
\!\left[
\psi^{\text{tree}}_{k_1 k_2 q_1 q_1'}
\right]
+
\int_{q_2 q_2'} P_{q_2 q_2'}
\,\mathrm{disc}_{q_2'}
\!\left[
\psi^{\text{tree}}_{k_1 k_2 q_2 q_2'}
\right]
\nonumber\\
&\quad+
\int_{q_1 q_1' q_2 q_2'}
P_{q_1 q_1'}P_{q_2 q_2'}
\,
\mathrm{disc}_{q_2'}
\!\left[
\psi_{k_1 q_1 q_2'}
\right]
\,
\mathrm{disc}_{q_1'}
\!\left[
\psi_{k_2 q_2 q_1'}
\right].
\end{align}
The key identity to arrive at \eqref{eq:2p1lCTT} from \eqref{eq:2p1l} is the decomposition of the Feynman bulk-bulk propagator as,
\begin{equation}\label{eq:GRdecompds}
G_p(\eta_1,\eta_2)
=
G_p^R(\eta_1,\eta_2)
+
2P_p\,K_p(\eta_1)\,
\mathrm{Im}[K_p(\eta_2)] .
\end{equation}
where,
\begin{equation}\label{eq:GRds}
    G_p^R(\eta_1, \eta_2,\vec{p}) = 2 P_p \mathrm{Im}\left[K_{p}(\eta_1)K^{*}_{p}(\eta_2)\right]\Theta(\eta_1-\eta_2),
\end{equation}
is the retarded propagator in dS.
Substituting \eqref{eq:GRdecompds} into \eqref{eq:2p1l}, the purely retarded contribution vanishes by causality \cite{AguiSalcedo:2023nds}.
Using the identity, 
$    i\,\mathrm{Im}[K_k]=\mathrm{disc}_k[K_k]$, one then obtains \eqref{eq:2p1lCTT}.

 In later sections, we work in(EA)dS using the continuation $\eta \rightarrow z=-i\eta$, where we will use "disc" operation rather than the "Im" one.

\subsection{Uplifting CTT from FTT via dressing}

In this section, we derive the cosmological tree theorem by appropriately dressing the Feynman tree theorem. This procedure systematically incorporates the effects of the curved background, allowing loop-level cosmological observables to be expressed entirely in terms of dressed tree-level data. In this way, the cosmological tree theorem emerges as a natural extension of its flat-space analogue, providing a unified framework for understanding loop dynamics in (EA)dS.

\subsubsection*{Conformally coupled $\phi^4$}
The relevant propagators for scalar theory in(EA)dS are \cite{Chowdhury:2025ohm},
\begin{align}
B_\nu(z,k) &=
- \sin(\pi \nu)\, z^{d/2}\,
\frac{2^{-\nu} k^\nu}{\Gamma(1+\nu)}\,
K_\nu(k z), \label{eq:btboundary} \\
\tilde{G}_\nu(z,z';k) &=
- \sin(\pi \nu)
\int dp\, \frac{p}{p^2 + k^2}\,
(z z')^{d/2}\,
J_\nu(p z)\, J_\nu(p z'),\label{eq:btb} \\
& = -2\sin(\pi \nu)\,(z z')^{d/2}\left\{
\Theta(z - z')\, K_\nu(kz)\, I_\nu(kz')
+ \Theta(z' - z)\, I_\nu(kz)\, K_\nu(kz')
\right\} \nonumber .
\end{align}
The functions $I_{\nu}$ and $K_{\nu}$ denote the modified Bessel functions of the first and second kind, respectively, while $J_{\nu}$ represents the Bessel function of the first kind. Furthermore, $B_{\nu}(z,\vec{k})$ corresponds to the bulk-to-boundary propagator, and $\tilde{G}_{\nu}(z,z';\vec{k})$ denotes the bulk-to-bulk propagator. For conformally coupled scalar fields, the parameter takes the values $\nu = \pm \tfrac{1}{2}$.
We can use the identity, 
\begin{equation}
    I_\nu(kz) = \frac{\Gamma(1+\nu)\, 2^\nu}{i \pi e^{i\pi\nu} \sin(\pi\nu)} k^{-\nu} z^{-d/2} (B_\nu(z, -k) - B_\nu(z, k)),
\end{equation}
and \eqref{eq:btboundary}, to rewrite \eqref{eq:btb} in terms of bulk-to-boundary propagator as follows,
\begin{align}\label{eq:GRdecompoEAdS}
    \tilde{G}_\nu(z,z';\vec{k}) &= \mathcal{F}  \left\{
\Theta(z - z')\, B_{\nu}(z,k)\, B_{\nu}(z',-k)
+ \Theta(z' - z)\, B_{\nu}(z,-k)\, B_{\nu}(z',k)-B_{\nu}( z,k)B_{\nu}(z',k)\right\}\,\nonumber\\
&  = \mathcal{F}  \Big\{
\Theta(z - z')\, \left(B_{\nu}(z,k)\, B_{\nu}(z',-k) - B_{\nu}(z,-k)\, B_{\nu}(z',k) \right) \nonumber
\\
&\qquad \qquad \qquad \qquad \qquad \qquad \qquad \qquad +  B_{\nu}(z,-k)\,B_{\nu}(z',k)-B_{\nu}(z,k)B_{\nu}( z',k)\Big\}\,\nonumber \\ 
& = \tilde{G}^R_\nu(z,z';\vec{k}) + \tilde{G}^{\mathrm{disc}}_{\nu}(z,z';\vec{k}) ,
\end{align}
where, $\mathcal{F}_{\nu}(k)$ is the overall factor,
\begin{equation}
    \mathcal{F}_{\nu}(k) = \frac{\Gamma(1+\nu)^2\, 2^{2\nu+1}}{i \pi e^{i\pi\nu} \sin(\pi\nu)} k^{-2\nu},
\end{equation}
and,
\begin{equation} \label{eq:GREAdS}
     \tilde{G}^R_\nu(z,z';\vec{k})= 2\mathcal{F}_{\nu}(k)\,
\mathrm{disc}_k
 [B_{\nu}(z,k)\, B_{\nu}(z',-k)]\Theta(z - z'),
\end{equation}
\begin{equation}\label{eq:Gdisc}
     \tilde{G}^{\mathrm{disc}}_{\nu}(z,z';\vec{k})  = -2\mathcal{F}_{\nu}(k)\,\mathrm{disc}_k[B_{\nu}(z,k)]B_{\nu}(z',k).
\end{equation}
Here, \eqref{eq:GREAdS}  is the(EA)dS version of \eqref{eq:GRds}\footnote{For a conformally coupled scalar, the bulk-to-boundary propagators are, $K_{k}(\eta) \approx e^{i k \eta}$ in de Sitter space and  $B_{\frac{1}{2}}(k z) \approx e^{-k z}$ in(EA)dS, with the two related by $\eta \rightarrow i z$. Under this continuation, one finds that   $\mathrm{Im}[K_{k}(\eta)K^*_{k}(\eta')]$ maps to $ \mathrm{disc}_k [B_{\frac{1}{2}}(k z) B_{\frac{1}{2}}(k z') ].$}. Let us now illustrate the uplifting mechanism of the Flat-space tree theorem to the cosmological tree theorem, with a simple example.

\subsubsection*{Six-point tree level Wave function coefficient}
Consider the s-channel exchange for six-point wave function coefficient. Using CTT we have the following diagrammatic  decomposition,
\begin{equation}
\begin{split}
    \begin{tikzpicture}[baseline = {(0,0.6)},thick,scale = 0.6]
\draw[dashed] (-3.5,2) -- (3.5,2);
\node at (-3,2.4) {$\mathbf{k_1}$};
\node at (-2,2.4) {$\mathbf{k_2}$};
\node at (-1,2.4) {$\mathbf{k_3}$};
\node at (1,2.4) {$\mathbf{k_4}$};
\node at (2,2.4) {$\mathbf{k_5}$};
\node at (3,2.4) {$\mathbf{k_6}$};
\coordinate (v1) at (-2,0);
\coordinate (v2) at (2,0);
\draw (-3,2) -- (v1);
\draw (-2,2) -- (v1);
\draw (-1,2) -- (v1);
\draw (1,2) -- (v2);
\draw (2,2) -- (v2);
\draw (3,2) -- (v2);
\draw (v1) -- (v2);
\node at (0,-0.5) {${p,\vec{s}}$};
\end{tikzpicture} 
=    \begin{tikzpicture}[baseline = {(0,0.6)},thick,scale = 0.6,
decoration={markings, mark=at position 0.5 with {\arrow{>}}}]
\draw[dashed] (-3.5,2) -- (3.5,2);
\node at (-3,2.4) {$\mathbf{k_1}$};
\node at (-2,2.4) {$\mathbf{k_2}$};
\node at (-1,2.4) {$\mathbf{k_3}$};
\node at (1,2.4) {$\mathbf{k_4}$};
\node at (2,2.4) {$\mathbf{k_5}$};
\node at (3,2.4) {$\mathbf{k_6}$};
\coordinate (v1) at (-2,0);
\coordinate (v2) at (2,0);
\draw (-3,2) -- (v1);
\draw (-2,2) -- (v1);
\draw (-1,2) -- (v1);
\draw (1,2) -- (v2);
\draw (2,2) -- (v2);
\draw (3,2) -- (v2);
\draw[postaction={decorate}] (v1) -- (v2);
\node at (0,-0.5) {${p,\vec{s}}$};
\end{tikzpicture} 
+  \int_{s\,s^\prime} P_{ss^\prime}\left(\begin{tikzpicture}[baseline = {(0,0.5)},thick,scale = 0.6,
decoration={markings, mark=at position 0.5 with {\arrow{>}}}]
\draw[dashed] (-3.5,2) -- (2.5,2);
\node at (-3.2,2.4) {$\mathbf{k_1}$};
\node at (-2.5,2.4) {$\mathbf{k_2}$};
\node at (-1.5,2.4) {$\mathbf{k_3}$};
\node at (-0.85,2.5) {$\mathbf{s'}$};
\node at (-0.2,2.4) {$\mathbf{s}$};
\node at (0.5,2.4) {$\mathbf{k_4}$};
\node at (1.3,2.4) {$\mathbf{k_5}$};
\node at (2,2.4) {$\mathbf{k_6}$};
\coordinate (v1) at (-2,0);
\coordinate (v2) at (1,0);
\draw (-3.2,2) -- (v1);
\draw (-2.5,2) -- (v1);
\draw (-1.7,2) -- (v1);
\draw[thick, double, draw = blue,style = dashed]   (-1,2) -- (v1);
\draw (-0.2,2) -- (v2);
\draw (0.5,2) -- (v2);
\draw (1.3,2) -- (v2);
\draw (2,2) -- (v2);
\end{tikzpicture}\right)
\end{split}
\end{equation}
where, the first diagram on RHS shows the contribution from $\tilde{G}^R$  and second from $\tilde{G}^{\mathrm{disc}}$ . Highlighted lines represent "disc" operation with respect to those momenta.

\

\noindent Below, we start with evaluating this diagram in(EA)dS using CTT and then via dressing the flat space FTT.
\subsubsection*{Direct calculation in(EA)dS using CTT}

The six-point tree-level wavefunction coefficient is written as,
\begin{align}
    \psi_{6,s} &= g^2\int\frac{dz_1}{z_1^4}\,\frac{dz_2}{z_2^4}\prod_{i=1}^3(z_1^\frac{3}{2}k_i^\frac{1}{2}K_\frac{1}{2}(k_iz_1))\prod_{j=4}^6(z_2^\frac{3}{2}k_j^\frac{1}{2}K_\frac{1}{2}(k_jz_2))\tilde{G}_{\frac{1}{2}} (z_1,z_2,\vec{s})\nonumber \\
    & = g^2 \int \frac{dz_1}{z_1} \frac{dz_2}{z_2}\,e^{-k_{123}z_1}\, e^{-k_{456}z_2} \,\tilde{G}_{\frac{1}{2}} (z_1,z_2,\vec{s}).
\end{align}
Using \eqref{eq:GRdecompoEAdS} we can write it as follows,
\begin{equation}
   \psi_{6,s} = g^2 \int \frac{dz_1}{z_1} \frac{dz_2}{z_2}\,e^{-k_{123}z_1}\, e^{-k_{456}z_2} \,\left(\tilde{G}^R_{\frac{1}{2}}(z_1,z_2;\vec{s}) + \tilde{G}^{\mathrm{disc}}_{\frac{1}{2}}(z_1,z_2;\vec{s}) \right) .
\end{equation}
We now proceed to evaluate the contributions from these two terms. Contribution from the first term
using definition of $\tilde{G}^R_{\frac{1}{2}}$ as in \eqref{eq:GREAdS} we get,
\begin{align}\label{eq:psi6s1}
    \psi^{(1)}_{6,s} &=g^2\,\int \frac{dz_1}{z_1} \frac{dz_2}{z_2}\,e^{-k_{123}z_1}\, e^{-k_{456}z_2} \,  \left(2\mathcal{F}_{\frac{1}{2}}(s)
\mathrm{disc}_s
 [B_{\frac{1}{2}}(z_1,s)\, B_{\frac{1}{2}}(z_2,-s)]\right)\Theta(z_1 - z_2)\, \nonumber \\
    & = \frac{g^2 }{s}\left( \frac{1}{E_T (E_L-2s)}-\frac{1}{E_T E_L} \right).
\end{align}

\noindent Next, the contribution from the second term
using definition of $\tilde{G}^{\mathrm{disc}}_{\frac{1}{2}}$ as given in \eqref{eq:Gdisc} we get,
\begin{align}\label{eq:psi6s2}
    \psi^{(2)}_{6,s} &= g^2 \int \frac{dz_1}{z_1} \frac{dz_2}{z_2}\,e^{-k_{123}z_1}\, e^{-k_{456}z_2} \,   \left(-2\mathcal{F}_{\frac{1}{2}}(s)\mathrm{disc}_s[B_{\frac{1}{2}}(z_1,s)]B_{\frac{1}{2}}( z_2,s)\right) \nonumber \\
    & = \frac{ g^2}{s} \left( \frac{1}{E_L E_R}-\frac{1}{E_R (E_L-2s)}  \right)\;.
\end{align}
Adding the contribution obtained in \eqref{eq:psi6s1} and \eqref{eq:psi6s2} we get,
\begin{align}\label{eq:psi6srdirect}
    \psi_{6,s} =   \psi^{(1)}_{6,s}+  \psi^{(2)}_{6,s} 
     =  \frac{ g^2}{s} \left(\frac{1}{E_L E_R} -\frac{1}{E_T E_L}-\frac{1}{E_T E_R}\right).
\end{align}
where, $E_T = k_1+k_2+k_3+k_4+k_5+k_6$,\quad $E_L = k_1+k_2+k_3+s$,\quad $E_R = k_4+k_5+k_6+s$, which correctly reproduces the wavefunction coefficient. We now revisit the same wavefunction coefficient from the viewpoint of dressing the flat-space tree theorem.

\subsubsection*{Via dressing FTT}

The flat space amplitude is given as,
\begin{align}\label{eq:ctt6pviadressing}
    \mathcal{A}^{\text{Tree}}_{6,s} &= (ig)^2 \int \frac{d^4k}{(2\pi)^4}\, \frac{i}{k^2}\, (2\pi)^4\delta^4(k_1+k_2+ k_3-k)\,(2\pi)^4\delta^4(k_4+k_5+k_6+k)\nonumber \\
    &= -\frac{g^2}{p^2 -s^2 +i\epsilon} \nonumber \\
    & = i g^2 \left( \Pi_R + \frac{\pi}{s}\delta (p+s) \right)\;,
\end{align}
where in the last line we have used \eqref{eq:GRdecompoflat}. We then apply the appropriate dressing factors to obtain the corresponding(EA)dS wavefunction coefficient. We begin by evaluating the contribution from the $\Pi_R$-term, which, upon dressing with the factors as in \eqref{eq:phi4dressing}, takes the form,
\begin{align}\label{eq:psi6sr}
     \psi^R_{6,s}  &= ig^2\int dz_1\, dz_2\, e^{-k_{123} z_1}\,e^{-k_{456} z_2}\, \int dp\, \Pi_R\, \mathrm{sin}(p z_1)\, \mathrm{sin}(p z_2) \nonumber \\
     &= ig^2\int  dz_1\, dz_2\, e^{-k_{123} z_1}\,e^{-k_{456} z_2}\, \Pi^d_R,
\end{align}
 where, $\Pi^d_R$ is the dressed retarded propagator as given in \eqref{eq:dressedpir}.
\noindent Recall that the dressing procedure involves a Wick rotating $s \to -is$. Under this transformation, the \eqref{eq:psi6sr} becomes,
 \begin{align}
    \psi^R_{6,s} &= \left(\frac{\pi g^2}{4 s}\right)\int dz_1\, dz_2\, e^{-k_{123} z_1}\,e^{-k_{456} z_2}\  
      \Bigg(\left(e^{- s (z_1+z_2)}-e^{ s (z_1+z_2)}\right)\nonumber \\
      &- \left(e^{- s (z_1-z_2)} -e^{ s (z_1-z_2)}\right)\theta(z_1-z_2)-\left(e^{- s (z_2-z_1)} -e^{ s (z_2-z_1)}\right)\theta(z_2-z_1)\Bigg)\;.
\end{align}
Performing $z_1,z_2$ integrals, we get,
\begin{align}\label{eq:psi6sR}
    \psi^R_{6,s} = \frac{\pi g^2}{4s}\left[\left(\frac{1}{E_L E_R}- \frac{1}{(E_L-2s) (E_R-2s)}\right)- \left( \frac{1}{E_T E_L}- \frac{1}{E_T (E_L -2 s)}\right) -\left(\frac{1}{E_T E_R}- \frac{1}{E_T (E_R -2 s)}\right)\right].
\end{align}
Next we look at the contribution coming from $\delta$-function term in \eqref{eq:ctt6pviadressing}. After dressing it takes the following form,
\begin{align}
    \psi^{\delta}_{6,s} &= ig^2\int dz_1\, dz_2\, e^{-k_{123} z_1}\,e^{-k_{456} z_2}\, \int dp\,\left(\frac{\pi}{s}\,\delta (p+s)\right)\, \mathrm{sin}(p z_1)\, \mathrm{sin}(p z_2)\nonumber \\
    & = \left(\frac{i\pi g^2}{s}\right)\int dz_1\, dz_2\, e^{-k_{123} z_1}\,e^{-k_{456} z_2}\, \mathrm{sin}(s z_1)\, \mathrm{sin}(s z_2).
\end{align}
Upon Wick rotation  $s\rightarrow -i s$, we get,
\begin{align}
 \psi^{\delta}_{6,s} &=  \frac{\pi g^2}{s}\int dz_1\, dz_2\, e^{-k_{123} z_1}\,e^{-k_{456} z_2}\, \mathrm{sinh}(s z_1)\, \mathrm{sinh}(s z_2) \nonumber \\
    &=  \frac{\pi g^2}{4s}\int dz_1\, dz_2\, e^{-k_{123} z_1}\,e^{-k_{456} z_2}\,\left(e^{-s(z_1+z_2)}-e^{-s(z_1-z_2)}-e^{s(z_1-z_2)}+e^{s(z_1+z_2)}\right).
\end{align}
Performing $z_i$ integrals, we get,
\begin{equation}\label{eq:psi6sdelta}
    \psi^{\delta}_{6,s} = \frac{\pi g^2}{4s}\left[\frac{1}{E_L E_R}-\frac{1}{E_L (E_R-2s)}-\frac{1}{ E_R (E_L-2s)}+\frac{1}{(E_L-2s)( E_R-2s)}\right].
\end{equation}
Now adding the contributions obtained in \eqref{eq:psi6sR} and \eqref{eq:psi6sdelta} we get,
\begin{equation}
    \psi_{6,s} = \frac{\pi g^2}{2s} \left(\frac{1}{E_L E_R} -\frac{1}{E_T E_L}-\frac{1}{E_T E_R}\right),
\end{equation}
which, up to an overall constant factor, agrees with \eqref{eq:psi6srdirect}.

Note that dressing the flat-space retarded propagator does not directly reproduce the EAdS retarded propagator \eqref{eq:GREAdS}, as discussed in the Appendix \ref{app:dressprop}. In particular, the uplift of the Feynman Tree Theorem involves non-trivial cancellations between the dressed retarded and  $\delta$-function contributions. These cancellations are essential for recovering the correct EAdS result. We will now investigate cases involving loop diagrams.

\subsubsection*{2-point 1-loop}
Consider the 2-point wave function coefficient at 1-loop
\begin{equation}\label{fig:2p1lphi4}
\begin{split}
\begin{tikzpicture}[baseline={(0,0.4)},scale = 0.9,thick]
    \draw[dashed, thick] (-1.1, 1) -- (1.1, 1);
    \draw[thick] (-0.8, 1) -- (-0.0, 0.25); \node[above] at (-0.6, 1) {$\mathbf{k}_1$};
    \draw[thick] (0.8, 1) -- (0.0, 0.24); \node[above] at (0.6, 1) {$\mathbf{k}_2$};
    \draw[thick, fill=white] (0, 0) circle (0.25);
    \node at (0, -0.45) {$\mathbf{q}$};
\end{tikzpicture}= \int_{\mathbf{q} \mathbf{q}^\prime} P_{\mathbf{q}\mathbf{q}^\prime}
\left(
\begin{tikzpicture}[baseline={(0,0.4)},scale = 0.8,thick]
    \draw[dashed, thick] (-1.5, 1) -- (1.5, 1);
    \draw[thick] (-1.2, 1) -- (0, 0); \node[above] at (-1.2, 1) {$\mathbf{k}_1$};
    \draw[thick] (-0.6, 1) -- (0, 0); \node[above] at (-0.6, 1) {$\mathbf{k}_2$};
    \draw[thick, double, draw = blue,style = dashed] (1.2, 1) -- (0, 0); \node[above] at (1.2, 1) {$\mathbf{q}^\prime$};
    \draw[thick] (0.6, 1) -- (0, 0); \node[above] at (0.6, 1) {$\mathbf{q}$};
\end{tikzpicture}
\right)
\end{split}
\end{equation}

\subsubsection*{Direct calculation in(EA)dS using CTT}

The wavefunction coefficient for diagram in  \eqref{fig:2p1lphi4} is written as,
\begin{align}
    \psi_{2}^{1-\mathrm{loop}} &= g\int d^3q\,\int \frac{dz}{z^4}\, \prod_{i=1}^2(z^\frac{3}{2}k_i^\frac{1}{2}K_\frac{1}{2}(k_iz))\,\tilde{G}_{\frac{1}{2}} (z_1,z_2,\vec{q})\nonumber \\
    & = g\int d^3q\,\int \frac{dz}{z^2}\, e^{-k_{12}z}\,\tilde{G}_{\frac{1}{2}} (z,z,\vec{q}).
\end{align}
Now, using \eqref{eq:GRdecompoEAdS} we can write,
\begin{equation}
   \psi_{2}^{1-\mathrm{loop}}= g\int d^3q\,\int \frac{dz}{z^2}\, e^{-k_{12}z}\,\left(\tilde{G}^R_{\frac{1}{2}}(z,z;\vec{q}) + \tilde{G}^{\mathrm{disc}}_{\frac{1}{2}}(z,z;\vec{q}) \right) .
\end{equation}
The term involving $\tilde{G}^{R}_{1/2}$ in the loop vanishes, as discussed earlier in Section \ref{sec:CTTds}. Consequently, only the second term contributes
\begin{align}\label{eq:2p1lredirect}
     \psi_{2}^{1-\mathrm{loop}} &= g \int d^3q\,\int \frac{dz}{z^2}\, e^{-k_{12}z}\,\left(-2\mathcal{F}_{\frac{1}{2}}(q)\,\mathrm{disc}_s[B_{\frac{1}{2}}(z,q)]\,B_{\frac{1}{2}}(z,q)\right) \nonumber \\
     & = g \int  \frac{d^3q}{q}\left(\frac{1}{k_{12}+2q}- \frac{1}{k_{12}}\right).
\end{align}
\subsubsection*{Via dressing FTT}

The flat space amplitude is written as,
\begin{align}
    \mathcal{A}^{1-\mathrm{loop}}_2 &= ig \int d^4L\,\frac{i}{L^2}\, \delta^3(\vec{k}_1+\vec{k}_2)\nonumber \\
    & =  ig \int d^3q\,dp\,\frac{1}{p^2-q^2}\, \delta^3(\vec{k}_1+\vec{k}_2)\nonumber \\
    & = g \int d^3q\,dp\,\left( \Pi_R + \frac{\pi}{q}\delta (p+q) \right),
\end{align}
where, $L = (p,\vec{q})$ is the loop four-momenta. In the last line we used \eqref{eq:GRdecompoflat}; we then apply the appropriate dressing factors to obtain the corresponding(EA)dS wavefunction coefficient.

We begin by evaluating the contribution from the $\Pi_R$-term, which, upon dressing with the factors as in \eqref{eq:phi4dressing}, takes the form,
\begin{align}
    \psi^{1-\mathrm{loop}}_{2(R)} = g \int d^3 q\, \int dz\, e^{-k_{12}z} \int dp\,  \Pi_R\, \mathrm{sin}^2(p z).
\end{align}
Performing the $p$-integral and Wick rotating $q \rightarrow -iq$ we get,
\begin{align}\label{R1}
      \psi^{1-\mathrm{loop}}_{2(R)} &= \frac{\pi g}{4} \int d^3 q\, \int dz\, e^{-k_{12}z} \frac{1}{q}\left(e^{-2qz}-e^{2qz} \right)\nonumber \\
    & =\frac{\pi g}{4} \int  \frac{d^3q}{q}\left( \frac{1}{k_{12}+2q}-\frac{1}{k_{12}-2q}\right).
\end{align}
Next, the contribution from $\delta$-function term, which, upon dressing, takes the following form,
\begin{align}
    \psi^{1-\mathrm{loop}}_{2(\delta)} &= g \int d^3 q\, \int dz\, e^{-k_{12}z} \int dp\, \left(\frac{\pi}{q}\,\delta (p+q)\right)\,\mathrm{sin}^2(p z)\nonumber \\
    &= \pi g \int \frac{d^3q}{q} \int dz\, e^{-k_{12}z} \mathrm{sin}^2(q z).
\end{align}
Upon Wick rotating $q \rightarrow -iq$ we get,
\begin{align}\label{D1}
      \psi^{1-\mathrm{loop}}_{2(\delta)} &=  \frac{ \pi g}{4} \int  \frac{d^3q}{q} \int dz\, e^{-k_{12}z} \left(e^{-2qz}+e^{2qz} -2\right)\nonumber \\
      & =  \frac{ \pi g}{4} \int  \frac{d^3q}{q} \left( \frac{1}{k_{12}+2q}+\frac{1}{k_{12}-2q}- \frac{2}{k_{12}}\right).
\end{align}
Finally, adding \eqref{R1} and \eqref{D1} we get
\begin{align}
    \psi_2^{1-\mathrm{loop}} &=  \psi^{1-\mathrm{loop}}_{2(R)} +\psi^{1-\mathrm{loop}}_{2(\delta)}\nonumber \\
    &=  \frac{ \pi g}{2} \int  \frac{d^3q}{q} \left(\frac{1}{k_{12}+2q}- \frac{1}{k_{12}}\right).
\end{align}
which, up to an overall constant factor, agrees with \eqref{eq:2p1lredirect}.

\subsubsection*{Four-point 1-loop}
We now consider the four-point wave function coefficient at the order of 1 loop.
\begin{equation}
\begin{split}
\begin{tikzpicture}[baseline={(0,0.6)},scale = 0.6,thick]
\draw[dashed] (-2.5,2) -- (2.5,2);
\node at (-2,2.4) {$\mathbf{k_1}$};
\node at (-1,2.4) {$\mathbf{k_2}$};
\node at (1,2.4) {$\mathbf{k_3}$};
\node at (2,2.4) {$\mathbf{k_4}$};
\coordinate (v1) at (-1.5,0);
\coordinate (v2) at (1.5,0);
\draw (-2,2) -- (v1);
\draw (-1,2) -- (v1);
\draw (1,2) -- (v2);
\draw (2,2) -- (v2);
\draw (v1) to[out=30,in=150] (v2);
\draw (v1) to[out=-30,in=-150] (v2);
\node at (0,0.8) {$\mathbf{q_1}$};
\node at (0,-0.8) {$\mathbf{q_2}$};
\end{tikzpicture}
&= \int_{\mathbf{q}_1 \mathbf{q}_1^\prime} P_{\mathbf{q}_1 \mathbf{q}_1^\prime}
\left(
\begin{tikzpicture}[baseline={(0,0.4)},scale = 0.8,thick]
    \draw[dashed, thick] (-2, 1) -- (2, 1);
    \draw[thick] (-1.8, 1) -- (-0.3, 0); \node[above] at (-1.8, 1) {$\mathbf{k}_1$};
      \draw[thick] (-1.2, 1) -- (-0.3, 0); \node[above] at (-1.2, 1) {$\mathbf{k}_2$};
    \draw[thick, double, draw = blue,style = dashed] (-0.6, 1) -- (-0.3, 0); \node[above] at (-0.6, 1) {$\mathbf{q}_1^\prime$};
    \draw (0.6, 1) -- (0.3, 0); \node[above] at (0.6, 1) {$\mathbf{q}_1$};
    \draw[thick] (1.2, 1) -- (0.3, 0); \node[above] at (1.2, 1) {$\mathbf{k}_3$};
     \draw[thick] (1.8, 1) -- (0.3, 0); \node[above] at (1.8, 1) {$\mathbf{k}_4$};
    \draw[thick] (-0.3, 0) -- (0.3, 0); \node[below] at (0, 0) {$\mathbf{q}_2$};
\end{tikzpicture}
\right)
+ \int_{\mathbf{q}_2 \mathbf{q}_2^\prime} P_{\mathbf{q}_2 \mathbf{q}_2^\prime}
\left(
\begin{tikzpicture}[baseline={(0,0.4)},scale = 0.8,thick]
    \draw[dashed, thick] (-2, 1) -- (2, 1);
    \draw[thick]  (-1.8, 1) -- (-0.3, 0); \node[above] at  (-1.8, 1) {$\mathbf{k}_1$};
        \draw[thick]  (-1.2, 1) -- (-0.3, 0); \node[above] at  (-1.2, 1) {$\mathbf{k}_2$};
    \draw[thick, double,  draw = blue,style = dashed] (-0.6, 1) -- (-0.3, 0); \node[above] at (-0.6, 1) {$\mathbf{q}_2^\prime$};
    \draw[thick] (0.6, 1) -- (0.3, 0); \node[above] at (0.6, 1) {$\mathbf{q}_2$};
    \draw[thick] (1.2, 1) -- (0.3, 0); \node[above] at (1.2, 1) {$\mathbf{k}_3$};
       \draw[thick] (1.8, 1) -- (0.3, 0); \node[above] at (1.8, 1) {$\mathbf{k}_4$};
    \draw[thick] (-0.3, 0) -- (0.3, 0); \node[below] at (0, 0) {$\mathbf{q}_1$};
\end{tikzpicture}
\right) \\
&\quad + \int_{\substack{\mathbf{q}_1, \mathbf{q}_2 \\ \mathbf{q}_1^\prime, \mathbf{q}_2^\prime}} P_{\mathbf{q}_1 \mathbf{q}_1^\prime} P_{\mathbf{q}_2 \mathbf{q}_2^\prime}
\left(
\begin{tikzpicture}[baseline={(0,0.4)},scale = 0.8,thick]
    \draw[dashed, thick] (-1.5, 1) -- (1.5, 1);
    \draw[thick] (-1.2, 1) -- (0, 0); \node[above] at (-1.2, 1) {$\mathbf{k}_1$};
    \draw[thick] (-0.6, 1) -- (0, 0); \node[above] at (-0.6, 1) {$\mathbf{k}_2$};
    \draw[thick, double, draw = blue,style = dashed] (1.2, 1) -- (0, 0); \node[above] at (1.2, 1) {$\mathbf{q}_2^\prime$};
    \draw[thick] (0.6, 1) -- (0, 0); \node[above] at (0.6, 1) {$\mathbf{q_1}$};
\end{tikzpicture}
\right)
\left(
\begin{tikzpicture}[baseline={(0,0.4)},scale = 0.8,thick]
    \draw[dashed, thick] (-1.5, 1) -- (1.5, 1);
    \draw[thick] (-1.2, 1) -- (0, 0); \node[above] at (-1.2, 1) {$\mathbf{k}_3$};
    \draw[thick] (-0.6, 1) -- (0, 0); \node[above] at (-0.6, 1) {$\mathbf{k}_4$};
    \draw[thick, double, draw = blue,style = dashed] (1.2, 1) -- (0, 0); \node[above] at (1.2, 1) {$\mathbf{q}_1^\prime$};
    \draw[thick] (0.6, 1) -- (0, 0); \node[above] at (0.6, 1) {$\mathbf{q}_2$};
\end{tikzpicture}
\right)
\end{split}
\end{equation}

\subsubsection*{Direct calculation in(EA)dS using CTT}
The wavefunction coefficient for above diagram is written as,
\begin{align}\label{eq:psi41loopinte}
 \psi^{1-\mathrm{loop}}_{4}
    &= g^2\int_{q_1\,q_2}\frac{dz_1}{z_1^4}\,\frac{dz_2}{z_2^4}\prod_{i=1}^2(z_1^\frac{3}{2}k_i^\frac{1}{2}K_\frac{1}{2}(k_iz_1))\prod_{j=3}^4(z_2^\frac{3}{2}k_j^\frac{1}{2}K_\frac{1}{2}(k_jz_2))\tilde{G}_{\frac{1}{2}} (z_1,z_2,\vec{q}_1) \tilde{G}_{\frac{1}{2}} (z_1,z_2,\vec{q}_2) \nonumber \\
    &\qquad \qquad \qquad \qquad \qquad \qquad \qquad \times \delta^3(\vec{k}_1 + \vec{k}_2+\vec{q}_1-\vec{q}_2)\delta^3(\vec{k}_3 + \vec{k}_4-\vec{q}_1+\vec{q}_2)\nonumber \\
    & = g^2 \int_{q_1\,q_2}\frac{dz_1}{z_1^2} \frac{dz_2}{z_2^2}\,e^{-k_{12}z_1}\, e^{-k_{34}z_2} \,\tilde{G}_{\frac{1}{2}} (z_1,z_2,\vec{q}_1) \tilde{G}_{\frac{1}{2}} (z_1,z_2,\vec{q}_2)\nonumber \\
    &\qquad \qquad \qquad \qquad \qquad \qquad \qquad \times \delta^3(\vec{k}_1 + \vec{k}_2+\vec{q}_1-\vec{q}_2)\delta^3(\vec{k}_3 + \vec{k}_4-\vec{q}_1+\vec{q}_2).
\end{align}
Now, using \eqref{eq:GRdecompoEAdS} we can write,
\begin{align}\label{eq:GRGREAdS}
    \tilde{G}_{\frac{1}{2}} (z_1,z_2,\vec{q}_1) \tilde{G}_{\frac{1}{2}} (z_1,z_2,\vec{q}_2) &= \tilde{G}^R_{\frac{1}{2}} (z_1,z_2,\vec{q}_1) \tilde{G}^R_{\frac{1}{2}} (z_1,z_2,\vec{q}_2) + \tilde{G}^R_{\frac{1}{2}} (z_1,z_2,\vec{q}_1) \tilde{G}^{\mathrm{disc}}_{\frac{1}{2}} (z_1,z_2,\vec{q}_2) \nonumber \\
    &+ \tilde{G}^{\mathrm{disc}}_{\frac{1}{2}} (z_1,z_2,\vec{q}_1) \tilde{G}^R_{\frac{1}{2}} (z_1,z_2,\vec{q}_2)+\tilde{G}^{\mathrm{disc}}_{\frac{1}{2}} (z_1,z_2,\vec{q}_1) \tilde{G}^{\mathrm{disc}}_{\frac{1}{2}} (z_1,z_2,\vec{q}_2) .
\end{align}
We follow the same procedure as outlined in Section \ref{sec:CTTds}. Terms containing only as product of $\tilde{G}^R_{\nu}$ in the loop are discarded, and \eqref{eq:GRdecompoEAdS} is used again to express the final result in terms of $\tilde{G}_{\nu}$. 

\noindent Next, we argue that one arrives at the same integrand from flat space via dressing.
\subsubsection*{Via dressing FTT}
The amplitude for this diagram is written as.
\begin{align}\label{eq:A41loop}
\mathcal{A}^{\text{1-loop}}_{4} &= (ig)^2 \int \frac{d^4L_1}{(2\pi)^4}\, \frac{d^4L_2}{(2\pi)^4}\,\frac{i}{L_1^2}\, \frac{i}{L_2^2}\, (2\pi)^4\delta^4(K_1+K_2+L_1-L_2)\nonumber \\
    &= -ig^2  \int \frac{d^3q_1}{(2\pi)^4}\, \frac{d^3q_2}{(2\pi)^4}\, dp_1\,dp_2\,\left(\frac{1}{p_1^2 -q_1^2 +i\epsilon}\right)\left( \frac{1}{p_2^2 -q_2^2 +i\epsilon}\right)(2\pi)^4\delta^4(K_1+K_2+L_1-L_2)\nonumber \\
    & = i g^2 \int_{q_1\,q_2}dp_1\,dp_2\left( \Pi_{R_1} + \frac{\pi}{q_1}\delta (p_1+q_1) \right)\left( \Pi_{R_2} + \frac{\pi}{q_2}\delta (p_2+q_2) \right)(2\pi)^3\delta^3(\vec{k}_1+\vec{k}_2+\vec{q}_1-\vec{q}_2),
\end{align}
where, $K_i = (k_i,\vec{k}_i)$, $L_i = (p,\vec{q}_i)$ and $\Pi_{R_i} \equiv \Pi_{R}(p_i,q_i)$. In the final line, we have used \eqref{eq:GRdecompoflat}. The integrand contains terms of the following form:
\begin{align}\label{eq:PiPi}
      \Pi_{F_1}\, \Pi_{F_2} = \Pi_{R_1}\, \Pi_{R_2} + \Pi_{R_1}\, \delta_2 + \Pi_{R_2}\, \delta_1 +\delta_1\,\delta_2,
\end{align}
where, $\delta_i \equiv \frac{\pi}{q_i}\delta(p_i + q_i)$. By dressing the \eqref{eq:A41loop} with the appropriate factors as in \eqref{eq:phi4dressing}, one can show that it matches \eqref{eq:GRGREAdS}. Details of the same are given in Appendix \ref{ap:calc}.

\subsubsection*{Conformally coupled $\phi^3$}
We now switch to $\phi^3$ theory of conformally coupled scalars.
\subsubsection*{Four-point tree-level}
Consider the s-channel tree-level exchange diagram.
\begin{equation}
\begin{split}
    \begin{tikzpicture}[baseline = {(0,0.6)},thick,scale = 0.6]
\draw[dashed] (-3.5,2) -- (3.5,2);
\node at (-3,2.4) {$\mathbf{k_1}$};
\node at (-1,2.4) {$\mathbf{k_2}$};
\node at (1,2.4) {$\mathbf{k_3}$};
\node at (3,2.4) {$\mathbf{k_4}$};
\coordinate (v1) at (-2,0);
\coordinate (v2) at (2,0);
\draw (-3,2) -- (v1);
\draw (-1,2) -- (v1);
\draw (1,2) -- (v2);
\draw (3,2) -- (v2);
\draw (v1) -- (v2);
\node at (0,-0.5) {${p,\vec{s}}$};
\end{tikzpicture} 
=    \begin{tikzpicture}[baseline = {(0,0.6)},thick,scale = 0.6,
decoration={markings, mark=at position 0.5 with {\arrow{>}}}]
\draw[dashed] (-3.5,2) -- (3.5,2);
\node at (-3,2.4) {$\mathbf{k_1}$};
\node at (-1,2.4) {$\mathbf{k_2}$};
\node at (1,2.4) {$\mathbf{k_3}$};
\node at (3,2.4) {$\mathbf{k_4}$};
\coordinate (v1) at (-2,0);
\coordinate (v2) at (2,0);
\draw (-3,2) -- (v1);
\draw (-1,2) -- (v1);
\draw (1,2) -- (v2);
\draw (3,2) -- (v2);
\draw[postaction={decorate}] (v1) -- (v2);
\node at (0,-0.5) {${p,\vec{s}}$};
\end{tikzpicture} 
+  \int_{s\,s^\prime} P_{ss^\prime}\left(\begin{tikzpicture}[baseline = {(0,0.4)},thick,scale = 0.6,
decoration={markings, mark=at position 0.5 with {\arrow{>}}}]
\draw[dashed] (-3.5,2) -- (2.5,2);
\node at (-3.2,2.4) {$\mathbf{k_1}$};
\node at (-2,2.4) {$\mathbf{k_2}$};
\node at (-0.85,2.5) {$\mathbf{s'}$};
\node at (-0.2,2.4) {$\mathbf{s}$};
\node at (1,2.4) {$\mathbf{k_3}$};
\node at (2,2.4) {$\mathbf{k_4}$};
\coordinate (v1) at (-2,0);
\coordinate (v2) at (1,0);
\draw (-3.2,2) -- (v1);
\draw (-2,2) -- (v1);
\draw[thick, double, draw = blue,style = dashed] (-1,2) -- (v1);
\draw  (-0.2,2) -- (v2);
\draw (2,2) -- (v2);
\draw (1,2) -- (v2);
\end{tikzpicture}\right)
\end{split}
\end{equation}
\subsubsection*{Direct calculations in(EA)dS using CTT}
The wavefunction coefficient is written as,
\begin{align}
    \psi_{4,s} &= g^2\int\frac{dz_1}{z_1^4}\,\frac{dz_2}{z_2^4}\prod_{i=1}^2(z_1^\frac{3}{2}k_i^\frac{1}{2}K_\frac{1}{2}(k_iz_1))\prod_{j=3}^4(z_2^\frac{3}{2}k_j^\frac{1}{2}K_\frac{1}{2}(k_jz_2))\tilde{G}_{\frac{1}{2}} (z_1,z_2,\vec{s})\nonumber \\
    & = g^2 \int \frac{dz_1}{z_1^2} \frac{dz_2}{z_2^2}\,e^{-k_{12}z_1}\, e^{-k_{34}z_2} \,\tilde{G}_{\frac{1}{2}} (z_1,z_2,\vec{s}).
\end{align}
Now, using \eqref{eq:GRdecompoEAdS} we can write,
\begin{equation}\label{treesplit}
   \psi_{4,s} = g^2 \int \frac{dz_1}{z_1^2} \frac{dz_2}{z_2^2}\,e^{-k_{12}z_1}\, e^{-k_{34}z_2} \,\left(\tilde{G}^R_{\frac{1}{2}}(z_1,z_2;\vec{s}) + \tilde{G}^{\mathrm{disc}}_{\frac{1}{2}}(z_1,z_2;\vec{s}) \right) .
\end{equation}
We now proceed to evaluate the contributions from these two terms separately. Contribution from the first term in \eqref{treesplit}
using definition of $\tilde{G}^R_{\frac{1}{2}}$ as in \eqref{eq:GREAdS} leads to
\begin{align}\label{eq:psi4s1}
    \psi^{(1)}_{4,s} &= g^2\,\int  \frac{dz_1}{z_1^2} \frac{dz_2}{z_2^2}\,e^{-k_{12}z_1}\, e^{-k_{34}z_2} \, \left(2\mathcal{F}_{\frac{1}{2}}(s)\,
\mathrm{disc}_s
 [B_{\frac{1}{2}}(z_1,s)\, B_{\frac{1}{2}}(z_2,-s)]\right)\Theta(z_1 - z_2)\, \nonumber \\
    & = -\frac{g^2}{s} \Bigg(\frac{\pi^{2}}{6} - \text{Li}_{2}\left(\frac{k_{12}-s}{k_{12}+k_{34}}\right) - \text{Li}_{2}\left(\frac{k_{34}-s}{k_{12}+k_{34}}\right)
+ \frac{1}{2}\ln^{2}\left(\frac{k_{12}+s}{k_{12}+k_{34}}\right) - \frac{1}{2}\ln^{2}\left(\frac{k_{12}-s}{k_{12}+k_{34}}\right)\nonumber \\ &\hspace{3cm}-\ln\left(\frac{k_{12}-s}{k_{12}+k_{34}}\right)\ln\left(\frac{k_{34}+s}{k_{12}+k_{34}}\right) + 2\ln(k_{12}+k_{34})\ln\left(\frac{k_{12}+s}{k_{12}-s}\right) \Bigg).
\end{align}
\noindent Next, the contribution from the second term in \eqref{treesplit}
using definition of $\tilde{G}^{\mathrm{disc}}_{\frac{1}{2}}$ as in \eqref{eq:Gdisc} yields,
\begin{align}\label{eq:psi4s2}
    \psi^{(2)}_{4,s} &= g^2\,\int  \frac{dz_1}{z_1^2} \frac{dz_2}{z_2^2}\,e^{-k_{12}z_1}\, e^{-k_{34}z_2} \, \left(-2\mathcal{F}_{\frac{1}{2}}(s)\,\mathrm{disc}_s[B_{\frac{1}{2}}(z_1,q)]B_{\frac{1}{2}}(z_2,q)\right) \nonumber \\
    & = \frac{g^2}{s} \Bigg(\frac{1}{2}\ln^{2}\left(\frac{k_{12}+s}{k_{12}+k_{34}}\right) - \frac{1}{2}\ln^{2}\left(\frac{k_{12}-s}{k_{12}+k_{34}}\right) + \ln\left(\frac{k_{12}+s}{k_{12}-s}\right) \ln\left(\frac{k_{34}+s}{k_{12}+k_{34}}\right) \nonumber \\ & \hspace{7cm} + 2\ln(k_{12}+k_{34})\ln\left(\frac{k_{12}+s}{k_{12}-s}\right) \Bigg).
\end{align}
The full wavefunction coefficient\footnote{Note that individual contributions contain dimensionful logarithms such as $\ln(k_{12}+k_{34})$. However, these terms cancel in the final expression, leaving a dimensionally consistent result.} is the sum of the contributions from \eqref{eq:psi4s1} and \eqref{eq:psi4s2},
\begin{align}\label{eq:psi4sdirect}
    \psi_{4,s} &= \psi^{(1)}_{4,s} +\psi^{(2)}_{4,s} \nonumber \\
    & = \frac{ g^2}{s}\left(\mathrm{Li}_2\left(\frac{k_{12}-s}{k_{12}+k_{34}}\right) + \mathrm{Li}_2\left(\frac{k_{34}-s}{k_{12}+k_{34}}\right)+ \mathrm{ln}\left(\frac{k_{12}+s}{k_{12}+k_{34}}\right)\,\mathrm{ln}\left(\frac{k_{34}+s}{k_{12}+k_{34}}\right)-\frac{\pi^2}{6}\right),
\end{align}
where, $\mathrm{Li}_2$ is the dilogarithm.

\subsubsection*{Via dressing FTT}

\noindent The flat space amplitude is given as,
\begin{align}\label{eq:ctt4pviadressing}
    \mathcal{A}^{\text{Tree}}_{4,s} &= (ig)^2 \int \frac{d^4k}{(2\pi)^4}\, \frac{i}{k^2}\, (2\pi)^4\delta^4(k_1+k_2-k)\,(2\pi)^4\delta^4(k_3+k_4+k)\nonumber \\
    &= -\frac{g^2}{p^2 -s^2 +i\epsilon} \nonumber \\
    & = i g^2 \left( \Pi_R + \frac{\pi}{s}\delta (p+s) \right).
\end{align}
In the last line we used \eqref{eq:GRdecompoflat}; we then apply the appropriate dressing factors to obtain the corresponding(EA)dS wavefunction coefficient.
\noindent We begin by evaluating the contribution from the $\Pi_R$-term, which, upon dressing with factors as in \eqref{eq:phi3dressing}, takes the form,
\begin{align}
     \psi^R_{4,s}  &= ig^2\int \frac{dz_1}{z_1}\, \frac{dz_2}{z_2}\, e^{-k_{12} z_1}\,e^{-k_{34} z_2}\, \int dp\, \Pi_R\, \mathrm{sin}(p z_1)\, \mathrm{sin}(p z_2) \nonumber \\
     &= ig^2\int \frac{dz_1}{z_1}\, \frac{dz_2}{z_2}\, e^{-k_{12} z_1}\,e^{-k_{34} z_2}\, \Pi^d_R,
\end{align}
 where, $\Pi^d_R$ is the dressed retarded propagator given in \eqref{eq:dressedpir}.
\noindent Recall that the dressing procedure involves a Wick rotation $s \to -is$. Under this transformation, the above equation becomes,
 \begin{align}
    \psi^R_{4,s} &= \left(\frac{\pi g^2}{4 s}\right)\int \frac{dz_1}{z_1}\, \frac{dz_2}{z_2}\, e^{-k_{12} z_1}\,e^{-k_{34} z_2}\  
      \Bigg(\left(e^{- s (z_1+z_2)}-e^{ s (z_1+z_2)}\right)\nonumber \\
      &- \left(e^{- s (z_1-z_2)} -e^{ s (z_1-z_2)}\right)\theta(z_1-z_2)-\left(e^{- s (z_2-z_1)} -e^{ s (z_2-z_1)}\right)\theta(z_2-z_1)\Bigg),
\end{align}
Upon performing $z_1,z_2$ integrals, we get,
\begin{align}\label{eq:psi4sR}
    \psi^R_{4,s} = \left(\frac{\pi g^2}{4 s}\right)&\Bigg(2\ln\left(\frac{k_{12}+s}{k_{12}+k_{34}}\right)\ln\left(\frac{k_{34}^{2}-s^{2}}{(k_{12}+k_{34})^{2}}\right) + \ln\left(\frac{k_{12}-s}{k_{12}+k_{34}}\right)\ln\left(\frac{k_{34}+s}{k_{34}-s}\right) \nonumber \\
    &\hspace{3cm}+ 2\text{Li}_{2}\left(\frac{k_{12}-s}{k_{12}+k_{34}}\right) + 2\text{Li}_{2}\left(\frac{k_{34}-s}{k_{12}+k_{34}}\right) - \frac{\pi^{2}}{3} \Bigg).
\end{align}
Next, we look at the contribution from $\delta$-function term which takes the following form after dressing,
\begin{align}
    \psi^{\delta}_{4,s} &= ig^2\int \frac{dz_1}{z_1}\, \frac{dz_2}{z_2}\, e^{-k_{12} z_1}\,e^{-k_{34} z_2}\, \int dp\,\left(\frac{\pi}{s}\,\delta (p+s)\right)\, \mathrm{sin}(p z_1)\, \mathrm{sin}(p z_2)\nonumber \\
    & = \left(\frac{i\pi g^2}{s}\right)\int \frac{dz_1}{z_1}\, \frac{dz_2}{z_2}\, e^{-k_{12} z_1}\,e^{-k_{34} z_2}\, \mathrm{sin}(s z_1)\, \mathrm{sin}(s z_2),
\end{align}
Upon Wick rotating  $s\rightarrow -i s$, we get,
\begin{align}
 \psi^{\delta}_{4,s} &=  \frac{\pi g^2}{s}\int \frac{dz_1}{z_1}\, \frac{dz_2}{z_2}\, e^{-k_{12} z_1}\,e^{-k_{34} z_2}\, \mathrm{sinh}(s z_1)\, \mathrm{sinh}(s z_2) \nonumber \\
    &=  \frac{\pi g^2}{4s}\int \frac{dz_1}{z_1}\, \frac{dz_2}{z_2}\, e^{-k_{12} z_1}\,e^{-k_{34} z_2}\,\left(e^{-s(z_1+z_2)}-e^{-s(z_1-z_2)}-e^{s(z_1-z_2)}+e^{s(z_1+z_2)}\right),
\end{align}
and after performing $z_1,z_2$ integrals, we finally get,
\begin{align}\label{eq:psi4sdelta}
    \psi^{\delta}_{4,s} = \frac{\pi g^2}{4s}&\Bigg(\mathrm{ln}\left(\frac{k_{12}-s}{k_{12}+s}\right)\,\mathrm{ln}\left(\frac{k_{34}-s}{k_{34}+s}\right)\Bigg).
\end{align}
Combining the contributions from \eqref{eq:psi4sR} and \eqref{eq:psi4sdelta} we get,
\begin{equation}
    \psi_{4,s} = \frac{\pi g^2}{2s}\left(\mathrm{Li}_2\left(\frac{k_{12}-s}{k_{12}+k_{34}}\right) + \mathrm{Li}_2\left(\frac{k_{34}-s}{k_{12}+k_{34}}\right)+ \mathrm{ln}\left(\frac{k_{12}+s}{k_{12}+k_{34}}\right)\,\mathrm{ln}\left(\frac{k_{34}+s}{k_{12}+k_{34}}\right)-\frac{\pi^2}{6}\right).
\end{equation}
which, up to an overall constant factor, agrees with \eqref{eq:psi4sdirect} obtained via direct calculation. We will now discuss uplifting of recursion relations in flat space to de Sitter.

\section{BCFW Recursion}\label{BCFW}
The remarkable successes in the study of scattering amplitudes in momentum space relied on several important results. These include (but are not exclusive to) the soft theorems \cite{Weinberg:1965nx, Low:1958sn}, colour-kinematics duality \cite{Bern:2008qj}, double-copy relations \cite{Bern:2010ue}, and recursion relations \cite{Britto:2005fq}. Given that the cosmological correlators can be obtained by dressing flat-space amplitudes, it is desirable to understand these features for the dS correlators. In this section, we extend the BCFW recursion relations to de Sitter (dS) space via the dressing mechanism.

The BCFW recursion \cite{Britto:2005fq} is an on-shell recursion method to obtain higher point amplitudes in terms of the lower point function. In this section, we will discuss how one can uplift this result to obtain its (EA)dS analogue\footnote{See also \cite{Albayrak:2018tam,Albayrak:2019yve}, where where higher-point gluuon and graviton AdS amplitudes were computed via bulk perturbation
theory in momentum-space. See \cite{Chu:2023pea} for Mellin-space recursion relations for tree-level gluon AdS amplitudes.} \cite{Raju:2010by}. Let us start with a brief review of the flat space BCFW recursion relations.

\subsection{BCFW in flat space}
In flat space, one obtains the BCFW recursion relations by performing complex deformations of external momenta and exploiting the analytic structure and factorization properties of amplitudes. This offers a highly efficient alternative to traditional perturbative methods.
\vspace{-15pt}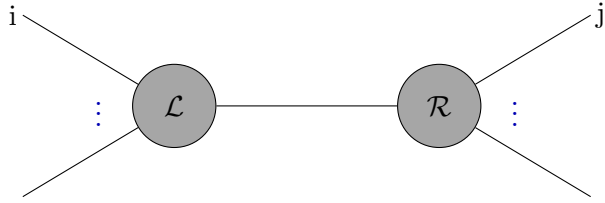
\begin{figure}[htbp]
    \centering
    \begin{tikzpicture}[
        blob/.style={circle, draw=black, fill=gray!70, minimum size=1.1cm, inner sep=0pt,scale = 1.0},
        labeltext/.style={text=blue!70!black}
    ]
        \begin{scope}[shift={(7.5,0)}] 
            \node[blob] (L) at (0, 0) {$\mathcal{L}$};
            \node[blob] (R) at (3.5, 0) {$\mathcal{R}$};

            \draw (L) -- (R);

            \draw (-2, 1.2) node[left, inner sep=2pt] {i} -- (L);
            \draw (-2, -1.2) -- (L);
            \node[labeltext] at (-1, 0) {$\vdots$};

            \draw (R) -- ++(2, 1.2) node[right, inner sep=2pt] {j};
            \draw (R) -- ++(2, -1.2);
            \node[labeltext] at (4.5, 0) {$\vdots$};

            \node[labeltext] at (1.75, -2.5){};
        \end{scope}

    \end{tikzpicture}
    \vspace{-40pt}\caption{A schematic representation of sub-amplitude factorization in recursion relations.}
\end{figure}

Consider an $n$-point tree-level amplitude $\mathcal{A}_n(k_1,\cdots,k_n,\epsilon_1,\cdots,\epsilon_n)$.
The BCFW shift is the following one-parameter momentum-conserving deformation on two legs leg i and j
\begin{equation}\label{bcfwdeform}
{k}_i\xrightarrow{}k_i^{\;'}={k}_i+c\;{q},\qquad {k}_j\xrightarrow{}k_j^{\;'}={k}_j-c\;{q},
\end{equation}
where ${q}$ is a null vector that preserves momentum conservation and on-shell condition
\begin{align}\label{bcfwnull}
{q}\cdot{q}={q}\cdot{k}_i={q}\cdot{k}_j=0,
\end{align}
The deformed amplitude is now a function of the parameter `$c$' \begin{align}\label{bcfwdeformamp}
\mathcal{A}_n(k_1,\cdots,k_n,\epsilon_1,\cdots,\epsilon_n)\rightarrow\mathcal{A}_n(k_1,\cdots,k_i^{\;'},\cdots,k_j^{\;'},\cdots,k_n,\epsilon_1,\cdots,\epsilon_i^{\;'},\cdots,\epsilon_j^{\;'},\cdots,\epsilon_n)
\end{align}    
where now $\epsilon^{\;'}_i\cdot k_i^{\;'}=\epsilon^{\;'}_j\cdot k_j^{\;'}=0$. The analytic structure of tree-level amplitude ensures that the poles occur only when an internal propagator goes on shell. Moreover, the amplitude factorizes into left and right sub-amplitudes, when the internal propagator goes on-shell at these poles. Under the BCFW shift, the propagator pole shifts from $P_{1\cdots i}=(k_1+\cdots+k_i)\rightarrow P_{1\cdots i}^{'}=(k_1+\cdots+k_i^{'})$. Consequently, the shifted amplitude has poles at
\begin{align}\label{bcfwpole}
c_i=-\frac{P_{1\cdots i}^2}{2q\cdot P_{1\cdots i}},
\end{align}
and the deformed amplitude factorizes as
\begin{align}\label{bcfwfact}
\mathcal{A}_n(c)\xrightarrow[c\to c_i]{}\;
\mathcal{A}_L(c_i)\,
\frac{1}{P_{1\cdots i}^{'\quad2}(c_i)}\,
\mathcal{A}_R(c_i).
\end{align}
The deformed amplitude in \eqref{bcfwdeformamp} is a complex meromorphic function that contains only simple poles in $c$. Thus, deforming the contour to infinity, and assuming that the amplitude dies off at large arcs, the physical amplitude can be written in terms of the sum over the residues at such poles. Evaluating the residue at these poles \eqref{bcfwpole}, and using the factorization \eqref{bcfwfact}, results in
\begin{align}\label{bcfwfinal}
&\mathcal{A}_n(k_1,\cdots,k_n,\epsilon_1,\cdots,\epsilon_n)\notag\\
&=\sum_{\textrm{P}}\mathcal{A}_L(k_1,\cdots,k_i^{'},\cdots,-P_{1\cdots i}^{'},\epsilon_1,\cdots,\epsilon_i^{'},\cdots,\epsilon_h)\,
\frac{1}{P_{1\cdots i}^{\quad2}}\,
\mathcal{A}_R(P_{1\cdots i}^{'},\cdots,k_j^{'},\cdots,k_n,\epsilon_h,\cdots,\epsilon_j^{'},\cdots,\epsilon_n),
\end{align}
where the $\sum_{\textrm{P}}$ is sum over all the possible factorizations, and the sum over polarizations $\epsilon_h$ is left implicit. Thus, one obtains an on-shell recursion relation where the n-point amplitude $\mathcal{A}_n$ is obtained by using lower point amplitudes i.e. $\mathcal{A}_L$ and $\mathcal{A}_R$ as building blocks.

We now turn to a brief discussion of the extension of BCFW like recursion relations obtained for (EA)dS correlators. In contrast to flat space, where these relations are formulated directly in terms of on-shell scattering amplitudes, their implementation in (EA)dS space requires adapting the complex momentum deformations to the boundary correlators.

\subsection{BCFW in (EA)dS space}

Motivated by their success in flat space, analogous analytic and factorization properties have been found for observables in (EA)dS \cite{Raju:2010by,Albayrak:2023jzl,Chowdhury:2024snc}. We briefly review these BCFW-like recursion relations obtained in the context of (EA)dS space below.

Consider the s-channel four-point (A)dS correlator for non-Abelian gauge fields in axial/temporal gauge
\begin{align}\label{ads4ptYM}
&\langle J(\vec{k}_1,\vec{\epsilon}_1)J(\vec{k}_2,\vec{\epsilon}_2)J(\vec{k}_3,\vec{\epsilon}_3)J(\vec{k}_4,\vec{\epsilon}_4)\rangle_{(s)}\notag\\
&=V_{12}^i\;V_{34}^j\;\pi_{ij}\int \frac{dz_1dz_2dp^2}{(z_1z_2)^{4}}\frac{(z_1 z_2)^{\frac{11}{2}}K_{\frac{1}{2}}(k_1z_1)K_{\frac{1}{2}}(k_2z_1)K_{\frac{1}{2}}(k_3z_2)K_{\frac{1}{2}}(k_4z_2)\;J_{\frac{1}{2}}(pz_1)J_{\frac{1}{2}}(pz_2)}{p^2+|\vec{k_1}+\vec{k_2}|^2},
\end{align}
where
\begin{align}\label{vertexYM}
V^k(k_i,k_j,\epsilon_i,\epsilon_j)=\epsilon_i\cdot\epsilon_j(k_i-k_j)^k+2(\epsilon_i\cdot k_j)\epsilon_j^k-2(\epsilon_j\cdot k_i)\epsilon_i^k.
\end{align}
One then performs the one-parameter shifts as mentioned in \eqref{bcfwdeform}, but restricted only to the boundary momenta. Note that in order to satisfy the constraints mentioned in \eqref{bcfwnull}, $\vec{q}$ must be complex. The key point is that the norm of momenta in the Bessel functions do not change under BCFW extension i.e. $|\vec{k}_i^{'}|=|\vec{k}|$, {by virtue of the choice of momentum shift in \eqref{bcfwdeform}}. However, for (EA)dS correlators, the pole now depend on the energy variable `$p$' due to energy non-conservation, unlike in the flat-space \eqref{bcfwpole}
\begin{align}\label{adspoleYM}
c_{s}(p)=-\frac{p^2+(\vec{k}_1+\vec{k}_2)^2}{2\vec{q}\cdot \vec{k}_2}. 
\end{align}
The residue at this poles turns out to be
\begin{align}\label{adsresidueYM}
\textrm{Res}_{c_s}&=\frac{1}{\vec{q}\cdot\vec{k_2}}\big(z_1^{\frac{3}{2}}K_{\frac{1}{2}}(k_1z_1)K_{\frac{1}{2}}(k_2z_1)J_{\frac{1}{2}}(pz_1)V_{12}^i\epsilon^{(s)}_i\big)\big(z_2^{\frac{3}{2}}K_{\frac{1}{2}}(k_3z_2)K_{\frac{1}{2}}(k_4z_2)J_{\frac{1}{2}}(pz_2)V_{34}^j\epsilon^{(s)}_j\big),
\end{align}
where we have used the completeness relation of the polarization tensors \eqref{eq:completeness}. Under these shifts, the exchange momenta across s-channel at the pole is precisely $p^2=|k_s^{'}|^2$. Therefore, the residue at the pole is precisely the product of the integrand for the three-point functions. Schematically, the recursion relation for the four-point, color-ordered (A)dS correlator takes the following form
\begin{align}\label{adsbcfwYM}
&\langle J(\vec{k}_1,\vec{\epsilon}_1)J(\vec{k}_2,\vec{\epsilon}_2)J(\vec{k}_3,\vec{\epsilon}_3)J(\vec{k}_4,\vec{\epsilon}_4)\rangle_{(s)}\notag\\
&=\int \frac{dp}{p^2+|\Vec{k_1}+\Vec{k_2}|^2}\langle J(\vec{k}_1^{\;'},\vec{\epsilon}_1^{\;'})J(\vec{k}_2,\vec{\epsilon}_2)J(-(\vec{k}_1^{\;'}+\vec{k}_2),\vec{\epsilon}_s)\rangle\;\langle J((\vec{k}_1^{\;'}+\vec{k}_2),\vec{\epsilon}_s)J(\vec{k}_3,\vec{\epsilon}_3)J(\vec{k}_4^{\;'},\vec{\epsilon}_4^{\;'})\rangle.
\end{align}
We now demonstrate how dressing the flat-space BCFW recursion relations, as given in \eqref{bcfwfinal}, with appropriate factors allows one to reproduce the corresponding recursion relations in (EA)dS.

\subsection{Uplifting BCFW recursion for (EA)dS correlators}

In this section, we obtain the (EA)dS analogue of the recursion relation \eqref{adsbcfwYM} by dressing flat space results as given in \eqref{bcfwfinal}. We start with the s-channel four-point amplitude for pure Yang-Mills theory. Using the flat space BCFW statement for this case \eqref{bcfwYM}, we have
\begin{align}\label{bcfwYM}
\mathcal{A}^{\textrm{YM}}_{4,s}(k,\epsilon)&=\frac{1}{s}\mathcal{A}^{\textrm{YM}}_3(k_1^{\;'},k_2,-(k_1^{\;'}+k_2),\epsilon_1^{\;'},\epsilon_2,\epsilon_s)\mathcal{A}^{\textrm{YM}}_3((k_1^{\;'}+k_2),k_3,k_4^{\;'},\epsilon_s,\epsilon_3,\epsilon_4^{\;'}),
\end{align}
where
\begin{align}\label{vertexYM}
\mathcal{A}_3^{\text{YM}}(k_1,k_2,k_3,\epsilon_1,\epsilon_2,\epsilon_3)=\epsilon_1\cdot\epsilon_2\;\epsilon_3\cdot k_1+\epsilon_2\cdot\epsilon_3\;\epsilon_1\cdot k_2+\epsilon_3\cdot\epsilon_1\;\epsilon_2\cdot k_3.
\end{align}
To obtain the corresponding statement for the (EA)dS correlator, we relax the energy conservation in the vertices and the propagator\footnote{We work in the axial/temporal gauge in order to have a clear comparison with the (EA)dS correlators.}, and then dress both the vertices with their respective auxiliary propagators as given in \eqref{eq:phi4dressing}. We restrict the BCFW shifts to adhere to \eqref{bcfwnull} for the boundary three-momenta, which requires us to work with complex $\vec{q}$ like in \cite{Raju:2010by}. Performing this procedure for the recursion relation given in \eqref{bcfwYM} yields,
\begin{align}\label{bcfwYMdressT}
&\langle J(\vec{k}_1,\vec{\epsilon}_1)J(\vec{k}_2,\vec{\epsilon}_2)J(\vec{k}_3,\vec{\epsilon}_3)J(\vec{k}_4,\vec{\epsilon}_4)\rangle_{(s)}\notag\\
&=\int \frac{dp}{p^2+|\Vec{k_1}+\Vec{k_2}|^2}\Bigg(\frac{k_{12}}{p^2+k_{12}^2}\mathcal{A}_3^{\textrm{YM}}(\vec{k}_1^{\;'},\vec{k}_2,-(\vec{k}_1^{\;'}+\vec{k}_2),\vec{\epsilon}_1^{\;'},\vec{\epsilon}_2,\vec{\epsilon}_s)\Bigg)\notag\\
&\qquad\qquad\qquad\qquad\quad\Bigg(\frac{k_{34}}{p^2+k_{34}^2}\mathcal{A}_3^{\textrm{YM}}((\vec{k}_1^{\;'}+\vec{k}_2),\vec{k}_3,\vec{k}_4^{\;'},\vec{\epsilon}_s,\vec{\epsilon}_3,\vec{\epsilon}_4^{\;'})\Bigg).
\end{align}
One observes that the first and second dressing factors in \eqref{bcfwYMdressT} correspond precisely to the dressings associated with the left and right sub-amplitudes, respectively\footnote{{These dressing factors are the result of performing the bulk-integral $z_i$ at each vertex. One can check that this matches with the form given in \cite{Raju:2010by} (which is presented in terms of $\frac{z^{\nu}}{z^{d+1}}KKJ$ integrand inside the $z$ integration), once the bulk integrals are carried out.}}. Thus, they result in the corresponding left and right sub-correlators
\begin{align}\label{eq:4p_rec}
&\langle J(\vec{k}_1,\vec{\epsilon}_1)J(\vec{k}_2,\vec{\epsilon}_2)J(\vec{k}_3,\vec{\epsilon}_3)J(\vec{k}_4,\vec{\epsilon}_4)\rangle_{(s)}\notag\\
&=\int \frac{dp}{p^2+|\Vec{k_1}+\Vec{k_2}|^2}\langle J(\vec{k}_1^{\;'},\vec{\epsilon}_1^{\;'})J(\vec{k}_2,\vec{\epsilon}_2)J(-(\vec{k}_1^{\;'}+\vec{k}_2),\vec{\epsilon}_s)\rangle\;\langle J((\vec{k}_1^{\;'}+\vec{k}_2),\vec{\epsilon}_s)J(\vec{k}_3,\vec{\epsilon}_3)J(\vec{k}_4^{\;'},\vec{\epsilon}_4^{\;'})\rangle,
\end{align}
which is the same as obtained in \cite{Raju:2010by}. Note that the shifted external momenta $\vec{k}_i^{\;'}$ is a function of the parameter $c_s$ through \eqref{bcfwdeform}, which depends on the energy variable $p$ (see \eqref{adspoleYM}. Thus, the RHS of \eqref{eq:4p_rec} depends on $p$ and renders a recursion at the integrand level, unlike in flat space.

Thus, we observe that the flat-space BCFW relation can be uplifted to its (EA)dS counterpart by dressing the shifted amplitudes with the same dressing as alluded in \eqref{eq:phi4dressing}. Notably, the BCFW shifts do not alter the arguments of the Bessel functions, since $|k^{\;'}|=|k|$. As the magnitudes are the relevant quantities in the dressing procedure, the dressing operation effectively bypasses the details of the BCFW algorithm.

A similar procedure can be carried out for graviton scattering by employing the dressing prescription given in \eqref{eq:gravitydressing}. Consider the BCFW statement for s-channel exchange of the graviton four-point function 
\begin{align}\label{bcfwGR}
\mathcal{A}^{\textrm{GR}}_{4,s}(k,\epsilon)&=\frac{1}{s}\mathcal{A}^{\textrm{GR}}_3(k_1^{\;'},k_2,-(k_1^{\;'}+k_2),\epsilon_1^{\;'},\epsilon_2,\epsilon_s)\mathcal{A}^{\textrm{GR}}_3((k_1^{\;'}+k_2),k_3,k_4^{\;'},\epsilon_s,\epsilon_3,\epsilon_4^{\;'}),
\end{align}
where the three-point graviton amplitude is related to three-point Yang-Mills amplitude via double-copy relation \cite{Bern:2010ue}
\begin{align}\label{doublecopy}
\mathcal{A}_3^{\text{GR}}(k_1,k_2,k_3,\epsilon_1,\epsilon_2,\epsilon_3)=\big(\mathcal{A}_3^{\text{YM}}(k_1,k_2,k_3,\epsilon_1,\epsilon_2,\epsilon_3)\big)^2.
\end{align}
Uplifting the flat-space BCFW relation for gravity \eqref{bcfwGR} by appropriately dressing it with \eqref{eq:gravitydressing}, results in (EA)dS recursion relation shown in \cite{Raju:2010by}

\begin{align}\label{adsbcfwgr}
&\langle T(\vec{k}_1,\vec{\epsilon}_1)T(\vec{k}_2,\vec{\epsilon}_2)T(\vec{k}_3,\vec{\epsilon}_3)T(\vec{k}_4,\vec{\epsilon}_4)\rangle_{(s)}\notag\\
&=\int \frac{dp}{p^2+|\Vec{k_1}+\Vec{k_2}|^2}\Big(\hat{\Delta}^T_h(k_1,k_2,p)
\mathcal{A}_3^{\textrm{GR}}(\vec{k}_1^{\;'},\vec{k}_2,-(\vec{k}_1^{\;'}+\vec{k}_2),\vec{\epsilon}_1^{\;'},\vec{\epsilon}_2,\vec{\epsilon}_s)\Big)\notag\\
&\qquad\qquad\qquad\qquad\quad\Big(\hat{\Delta}^T_h(k_1,k_2,p)\mathcal{A}_3^{\textrm{GR}}((\vec{k}_1^{\;'}+\vec{k}_2),\vec{k}_3,\vec{k}_4^{\;'},\vec{\epsilon}_s,\vec{\epsilon}_3,\vec{\epsilon}_4^{\;'})\Big)\notag\\
&=\int \frac{dp}{p^2+|\Vec{k_1}+\Vec{k_2}|^2}\langle T(\vec{k}_1^{\;'},\vec{\epsilon}_1^{\;'})T(\vec{k}_2,\vec{\epsilon}_2)T(-(\vec{k}_1^{\;'}+\vec{k}_2),\vec{\epsilon}_s)\rangle\;\langle T((\vec{k}_1^{\;'}+\vec{k}_2),\vec{\epsilon}_s)T(\vec{k}_3,\vec{\epsilon}_3)T(\vec{k}_4^{\;'},\vec{\epsilon}_4^{\;'})\rangle.
\end{align}
Thus we observe that one can recursively obtain the higher point gluons and gravitons correlators in (EA)dS by suitably dressing the flat-space BCFW relations. While we have shown this result explicitly at the level of four-point functions, this recursion process can be very easily generalized to any tree-level, higher-point function.

The BCFW recursion relations exploit the analytic structure of amplitudes, but an equally powerful set of constraints arises from their behaviour in singular kinematic limits. In particular, soft theorems capture the universal structure of amplitudes when one of the external momenta is taken to be soft. Let us now move on to discuss soft theorems in flat space and how one can uplift them to get soft theorems for (EA)dS correlators.

\section{Soft Theorems}\label{Soft}

The soft theorem \cite{Weinberg:1965nx} is one of the pivotal results in the study of scattering amplitudes. It is a simple consequence of gauge invariance, wherein an amplitude factorizes into a product of a lower-point amplitude and a universal factor, when one of the external legs is taken to be soft \cite{Bern:2014vva}. In this section, we will discuss how one can uplift this result for its (EA)dS counterpart. Let us quickly review soft theorems in flat space.

\subsection{Soft limit in flat space}
Consider an n-point color ordered amplitude in Yang-Mills theory where the $n^\textrm{{th}}$ leg is taken soft i.e. $k_n\rightarrow 0$. The amplitude admits the following series expansion in terms of soft momenta,
\begin{align}\label{softexp}
\lim_{k_n\rightarrow0}\mathcal{A}_{n}(k_1,\cdots,k_n)=\Big(S_n^{(0)}+S_n^{(1)}+\cdots\Big)\mathcal{A}_{n}(k_1,\cdots,k_{n-1}),
\end{align}
where $S_n^{(i)}$'s are ``universal soft factors" depending on the momentum taken to be soft. The leading order soft factor is a spin-independent universal structure $S_n^{(0)}$ which comes from the standard soft pole and is given as,
\begin{equation}\label{leadsoft}
S_n^{(0)}=\sum_{i=1}^{n-1}\frac{k_i\cdot\epsilon_n}{k_i\cdot k_n}T_i,
\end{equation}
while the sub-leading contribution $S_n^{(1)}$ encodes the spin of the hard leg (here, $i^\textrm{th}$ leg)\footnote{The hard leg is defined as the external leg to which soft leg is attached and whose momenta is large compared to the soft leg.} and takes the following form,
\begin{align}\label{subsoft}
S_n^{(1)}=\sum_{i=1}^{n-1}\epsilon_n^\mu k_n^\nu\frac{J_i^{\mu\nu}}{k_i\cdot k_n}T_i,
\end{align}
where $J_i^{\mu\nu}$ is given as
\begin{align}\label{angmom}
J_i^{\mu\nu}=\Big(\epsilon_i^\mu\frac{\partial}{\partial\epsilon_i^\nu}-\epsilon_i^\nu\frac{\partial}{\partial\epsilon_i^\mu}\Big)+\Big(k_i^\mu\frac{\partial}{\partial k_i^\nu}-k_i^\nu\frac{\partial}{\partial k_i^\mu}\Big),
\end{align}
and $T_i$ are the color generator associated with $i^\textrm{th}$ particle.

Given their universality and model-independent nature, soft theorems are expected to provide powerful constraints on observables in (EA)dS/dS as well. In particular, soft limits can probe the infrared structure of cosmological correlators and potentially reveal underlying symmetries and consistency conditions. This makes them a promising tool for organizing and constraining (EA)dS observables beyond explicit perturbative calculations. Let us quickly look at the soft limits for (EA)dS correlators.

\subsection{Soft limit in (EA)dS space}
Soft limits in (EA)dS correspond to taking one external momentum of a correlator to be small, often leading to simplified and potentially universal structures analogous to flat-space soft theorems. These limits provide insight into the infrared structure and symmetry constraints of (EA)dS observables. For example, the soft limit of three-point correlator for conserved currents and stress tensor are shown to take the following form \cite{Maldacena:2011nz}
\begin{align}\label{adssoft}
\lim_{\vec{k}_3\rightarrow0}\langle J(k_1)J(k_2)J(k_3)\rangle&=-\frac{\epsilon_3\cdot k_2}{2k_2}\partial_{k_2}\langle J(k_1)J(k_2)\rangle,\notag\\
\lim_{\vec{k}_3\rightarrow0}\langle T(k_1)T(k_2)T(k_3)\rangle&=-\frac{(\epsilon_3\cdot k_2)^2}{2k_2}\partial_{k_2}\langle T(k_1)T(k_2)\rangle.
\end{align}
A similar universal structure has been found for higher-point (EA)dS correlators, albeit only at the leading order \cite{Albayrak:2024ddg,Chowdhury:2024wwe,Chowdhury:2024snc,Mei:2025jko}. The four-point function for Yang-Mills theory for s-channel exchange is given as
\begin{align}\label{adssoft4ptYM}
&\lim_{\vec{k}_4 \to 0}\langle J(\vec{k}_1,\vec{\epsilon}_1)J(\vec{k}_2,\vec{\epsilon}_2)J(\vec{k}_3,\vec{\epsilon}_3)J(\vec{k}_4,\vec{\epsilon}_4)\rangle\notag\\
&=\frac{\vec{\epsilon}_4\cdot \vec{k}_3}{k_3}\partial_{k_3}\langle J(\vec{k}_1,\vec{\epsilon}_1)J(\vec{k}_2,\vec{\epsilon}_2)J(\vec{k}_3,\vec{\epsilon}_3)\rangle-\frac{\vec{\epsilon}_4\cdot \vec{k}_3}{k_3^2}\vec{k}_3\cdot\partial_{\vec{\epsilon}_3}\langle J(\vec{k}_1,\vec{\epsilon}_1)J(\vec{k}_2,\vec{\epsilon}_2)J(\vec{k}_3,\vec{\epsilon}_3)\rangle.
\end{align}
This can be written in a more compact form \cite{Chowdhury:2024wwe}
\begin{align}\label{adssoft4ptYMfinal}
&\lim_{\vec{k}_4 \to 0}\langle J(\vec{k}_1,\vec{\epsilon}_1)J(\vec{k}_2,\vec{\epsilon}_2)J(\vec{k}_3,\vec{\epsilon}_3)J(\vec{k}_4,\vec{\epsilon}_4)\rangle=\frac{\epsilon_4\cdot\partial_{k_3}}{2}\langle J(\vec{k}_1,\vec{\epsilon}_1)J(\vec{k}_2,\vec{\epsilon}_2)J(\vec{k}_3,\vec{\epsilon}_3)\rangle.
\end{align}
In particular, a similar structure is observed for the four-point graviton correlator too \cite{Chowdhury:2024wwe}
\begin{align}\label{adssoft4ptGRfinal}
&\lim_{\vec{k}_4 \to 0}\langle T(\vec{k}_1,\vec{\epsilon}_1)T(\vec{k}_2,\vec{\epsilon}_2)T(\vec{k}_3,\vec{\epsilon}_3)T(\vec{k}_4,\vec{\epsilon}_4)\rangle=\frac{\epsilon_4^{ij}k_{3i}\partial_{k_{3j}}}{2}\langle T(\vec{k}_1,\vec{\epsilon}_1)T(\vec{k}_2,\vec{\epsilon}_2)T(\vec{k}_3,\vec{\epsilon}_3)\rangle.
\end{align}
Thus the leading soft limit of (EA)dS correlators have some universal feature, much like its flat space counterpart. Our motivation is to investigate whether dressing rules can be employed to derive a universal sub-leading soft factor for (EA)dS correlators, in analogy with the known universality of sub-leading soft limits in flat space. 

Let us start with uplift leading soft limits for the correlators by applying appropriate dressing to the flat-space soft limit.

\subsection{Uplifting leading soft limit for (EA)dS correlators}
We start with the case for scalar QED for simplicity. Consider the four-point correlator $\langle JOOJ\rangle$ in the s-channel with the scalar exchange\footnote{{We restrict our analysis to class I diagrams (see \cite{Chowdhury:2024wwe} for the classification) i.e. the exchange diagrams in this context, which are the sole contributors to (sub-)leading soft theorem in flat-space. However in AdS, the contact diagram also contributes to (sub-)leading soft-limit \cite{Chowdhury:2024wwe}. Its contribution to the soft-limits can be obtained by appropriate soft dressing of the contact flat-space amplitude. However, in this paper, our focus is only on the exchange diagrams (class I).}}
\begin{align}\label{ads4ptsqeds}
\langle J(\vec{k}_1)O(\vec{k}_2)O(\vec{k}_3)J(\vec{k}_4)\rangle=\frac{\vec{\epsilon}_1\cdot \vec{k}_2\; \vec{\epsilon}_4\cdot \vec{k}_3}{(k_1+k_2+k_3+k_4)(k_1+k_2+|\vec{k}_3+\vec{k}_4|)(k_3+k_4+|\vec{k}_3+\vec{k}_4|)}.
\end{align}
Taking a soft limit with respect to $\vec{k}_4$ results in
\begin{align}
\lim_{\vec{k}_4 \to 0}\langle J(\vec{k}_1)O(\vec{k}_2)O(\vec{k}_3)J(\vec{k}_4)\rangle=\frac{\vec{\epsilon}_4\cdot \vec{k}_3}{2k_3}\frac{\vec{\epsilon}_1\cdot \vec{k}_2}{(k_1+k_2+k_3)^2},
\end{align}
which takes a form reminiscent of the leading soft limit of its flat-space counterpart
\begin{align}\label{adssoft4ptsqeds}
\lim_{\vec{k}_4 \to 0}\langle J(\vec{k}_1)O(\vec{k}_2)O(\vec{k}_3)J(\vec{k}_4)\rangle=\frac{\vec{\epsilon}_4\cdot \vec{k}_3}{2k_3}\partial_{k_3}\langle  J(\vec{k}_1)O(\vec{k}_2)O(\vec{k}_3)\rangle.
\end{align}
Here, the pre-factor has the soft pole replaced with energy of the hard leg ie. $k_3$, multiplied by the three point correlator $\langle JOO\rangle$ with a higher order total energy pole.
  
We will now obtain this result by appropriately dressing the flat-space soft theorem. Consider the four-point amplitude for s-channel scalar exchange in scalar QED
\begin{align}\label{sqeds}
\mathcal{A}_4(k_1,k_2,k_3,k_4,\epsilon_1,\epsilon_4)=\frac{2\;\epsilon_4\cdot k_3\;{\epsilon_1\cdot k_2}}{k_4\cdot k_3}.
\end{align}
The leading soft limit in this case is simply
\begin{align}\label{softsqeds}
\lim_{k_4 \to 0}\mathcal{A}_4(k_1,k_2,k_3,k_4,\epsilon_1,\epsilon_4)=\frac{\epsilon_4\cdot k_3}{k_4\cdot k_3}\mathcal{A}_3(k_1,k_2,k_3,\epsilon_1),
\end{align}
where $\mathcal{A}_3(k_1,k_2,k_3,\epsilon_1)=\epsilon_1\cdot(k_2-k_3)$. We now relax the energy conservation in the propagator and dress \eqref{softsqeds} with appropriate dressing factors \eqref{eq:phi4dressing} at each vertex. However, since leg-4 is taken to be soft ($k_4 = 0$), the corresponding modified factors will be referred to as “soft” dressing factors.
 The integrand after this dressing takes the following form
\begin{align}\label{dresssoftsqeds}
\lim_{\vec{k}_4 \to 0}\langle J(\vec{k}_1)O(\vec{k}_2)O(\vec{k}_3)J(\vec{k}_4)\rangle
&=\vec{\epsilon}_4\cdot \vec{k}_3\;\vec{\epsilon}_1\cdot (\vec{k}_2-\vec{k}_3)\int \frac{dp}{p^2+|\Vec{k_1}+\Vec{k_2}|^2}\frac{k_{12}}{p^2+k_{12}^2}\frac{k_{3}}{p^2+k_{3}^2}.
\end{align}
where we take leg-4 to be soft in the dressing. Performing the energy integral results in
\begin{align}\label{adssoftsqeds}
\lim_{\vec{k}_4 \to 0}\langle J(\vec{k}_1)O(\vec{k}_2)O(\vec{k}_3)J(\vec{k}_4)\rangle
&=\frac{\vec{\epsilon}_4\cdot \vec{k}_3}{k_3(k_1+k_2+k_3)}\Big(\frac{1}{k_3}+\frac{1}{k_1+k_2+k_3}\Big)\;\vec{\epsilon}_1\cdot (\vec{k}_2-\vec{k}_3),
\end{align}
which matches the soft limit \eqref{adssoft4ptsqeds} up to a semi-local term
\begin{align}
\lim_{\vec{k}_4 \to 0}\langle J(\vec{k}_1)O(\vec{k}_2)O(\vec{k}_3)J(\vec{k}_4)\rangle
&=\frac{\vec{\epsilon}_4\cdot \vec{k}_3}{k_3(k_1+k_2+k_3)}\langle J(\vec{k}_1)O(\vec{k}_2)O(\vec{k}_3)\rangle.
\end{align}
Thus the leading soft limit of (EA)dS correlators can be obtained by `soft' dressing of the flat-space leading soft theorem.

\noindent A similar statement holds for Yang-Mills theory as well\footnote{{Working in the axial/temporal gauge, we begin by allowing for both: the transverse and longitudinal propagators in the exchange diagram. This peculiar choice of gauge becomes evident when studying wavefunction coefficients or correlators in cosmology. The longitudinal mode is essential for the correct late-time boundary condition for spinning fields \cite{Chowdhury:2025nnk}.}}
\begin{align}\label{dresssoftYM}
&\lim_{\vec{k}_4 \to 0}\langle J(\vec{k}_1,\vec{\epsilon}_1)J(\vec{k}_2,\vec{\epsilon}_2)J(\vec{k}_3,\vec{\epsilon}_3)J(\vec{k}_4,\vec{\epsilon}_4)\rangle\notag\\
&=\frac{\vec{\epsilon}_4\cdot \vec{k}_3}{k_3}\int \frac{dp}{p^2+|\Vec{k_1}+\Vec{k_2}|^2}\frac{k_{12}}{p^2+k_{12}^2}\frac{k_{3}}{p^2+k_{3}^2}\mathcal{A}_3(\vec{k}_1,\vec{k}_2,\vec{k}_3,\vec{\epsilon}_1,\vec{\epsilon}_2,\vec{\epsilon}_3)\notag\\
&+(k_1^2-k_2^2)(\vec{\epsilon}_1\cdot\vec{\epsilon}_2)(\vec{\epsilon}_3\cdot\vec{\epsilon}_4)\int \frac{dp}{p^2}\frac{k_{12}}{p^2+k_{12}^2}\frac{k_{3}}{p^2+k_{3}^2},
\end{align}
where the first and second  line in the RHS of \eqref{dresssoftYM} are the terms from transverse and longitudinal parts respectively. Now, performing the energy integral results in
\begin{align}\label{adssoft4ptYMlead}
&\lim_{\vec{k}_4 \to 0}\langle J(\vec{k}_1,\vec{\epsilon}_1)J(\vec{k}_2,\vec{\epsilon}_2)J(\vec{k}_3,\vec{\epsilon}_3)J(\vec{k}_4,\vec{\epsilon}_4)\rangle\notag\\
&=\frac{\vec{\epsilon}_4\cdot \vec{k}_3(\vec{\epsilon}_1\cdot\vec{\epsilon}_2\; \vec{k}_3\cdot\vec{\epsilon}_1+\textrm{perms})}{k_3(k_1+k_2+k_3)}\Big(\frac{1}{k_3}+\frac{1}{k_1+k_2+k_3}\Big)+\frac{(k_1-k_2)(\vec{\epsilon}_1\cdot\vec{\epsilon}_2)(\vec{\epsilon}_3\cdot\vec{\epsilon}_4)}{2k_3(k_1+k_2+k_3)},
\end{align}
which matches the leading soft limit result of four-point correlator in Yang-Mills theory \eqref{adssoft4ptYM} upto a semi-local term. Let us reiterate the soft limit for the (EA)dS correlator is
\begin{align}
&\lim_{\vec{k}_4 \to 0}\langle J(\vec{k}_1,\vec{\epsilon}_1)J(\vec{k}_2,\vec{\epsilon}_2)J(\vec{k}_3,\vec{\epsilon}_3)J(\vec{k}_4,\vec{\epsilon}_4)\rangle\notag\\
&=\frac{\vec{\epsilon}_4\cdot \vec{k}_3}{k_3(k_1+k_2+k_3)}\langle J(\vec{k}_1,\vec{\epsilon}_1)J(\vec{k}_2,\vec{\epsilon}_2)J(\vec{k}_3,\vec{\epsilon}_3)\rangle+\frac{(k_1-k_2)(\vec{\epsilon}_1\cdot\vec{\epsilon}_2)(\vec{\epsilon}_3\cdot\vec{\epsilon}_4)}{2k_3(k_1+k_2+k_3)}.
\end{align}
It was observed in \cite{Chowdhury:2024wwe}, that while the first term is a direct analogue of Weinberg soft theorem (with the soft pole now being replaced by an energy derivative with respect to the hard leg), the second term can be interpreted as a polarization-dependent derivative on the lower-point function
\begin{align}
&\lim_{\vec{k}_4 \to 0}\langle J(\vec{k}_1,\vec{\epsilon}_1)J(\vec{k}_2,\vec{\epsilon}_2)J(\vec{k}_3,\vec{\epsilon}_3)J(\vec{k}_4,\vec{\epsilon}_4)\rangle\notag\\
&=\frac{\vec{\epsilon}_4\cdot \vec{k}_3}{k_3}\partial_{k_3}\langle J(\vec{k}_1,\vec{\epsilon}_1)J(\vec{k}_2,\vec{\epsilon}_2)J(\vec{k}_3,\vec{\epsilon}_3)\rangle-\frac{\vec{\epsilon}_4\cdot \vec{k}_3}{k_3^2}\vec{k}_3\cdot\partial_{\vec{\epsilon}_3}\langle J(\vec{k}_1,\vec{\epsilon}_1)J(\vec{k}_2,\vec{\epsilon}_2)J(\vec{k}_3,\vec{\epsilon}_3)\rangle.
\end{align}
Hence, the leading soft limit of (EA)dS correlator encodes both: a universal factor and a spin-dependent part. Our analysis clearly shows that the latter stems from the longitudinal mode of the gluon exchange, presenting a clear source for the spin-dependent part in the leading soft limit of the correlator. Thus we observe that the leading soft limit for (EA)dS correlators can be obtained by dressing the leading soft limit of flat-space amplitudes with `soft' dressing factors\footnote{This is true upto the off-shell contributions from longitudinal propagator, which do not have an on-shell analogue for amplitudes.}.

We now discuss the procedure for obtaining the sub-leading soft limit of (EA)dS correlators, by uplifting the soft theorem in flat space.

\subsection{Uplifting sub-leading soft limit for (EA)dS correlators}
The subleading soft limit of (EA)dS correlators presents certain subtleties that require careful analysis. While the leading soft limit came unambiguously from the product of flat-space leading soft theorem with soft dressing, and the contribution from the longitudinal propagator, we find that the subleading contribution can arise from either of the following sources,
\begin{enumerate}
\item Sub-leading soft dressing of the leading soft theorem.
\item Leading soft dressing of the sub-leading soft theorem.
\item Off-shell parts and contributions from longitudinal propagator.
\end{enumerate}
In order to get this clear segregation, we approach the problem in a way that closely mimics the one in \cite{Bern:2014vva} for (sub)leading soft limit for the amplitudes. Let us illustrate this for the case of scalar QED. Consider the t-channel diagram for $\langle JOOJ\rangle$ amplitude
\begin{align}\label{eq:sqed}
\mathcal{A}_4^{(t)}(k_1,\epsilon_1,k_2,k_3,k_4,\epsilon_4)=\frac{V_{14}^{\mu}\Pi_{\mu\nu}(k_2^\nu-k_3^\nu)}{2k_1\cdot k_4}.
\end{align}
The soft expansion of this amplitude about $k_4$ is
\begin{align}\label{tsoft}
\lim_{k_4\rightarrow0}\mathcal{A}_4^{(t)}(k_1,\epsilon_1,k_2,k_3,k_4,\epsilon_4)=\frac{2k_1\cdot \epsilon_4 (k_2-k_3)\cdot \epsilon_1}{k_1\cdot k_4}+
\frac{
((k_2 - k_3)\cdot \epsilon_4)(\epsilon_1 \cdot k_4)
-
((k_2 - k_3)\cdot k_4)(\epsilon_1 \cdot \epsilon_4)
}{k_1\cdot k_4},
\end{align}
where the first term is the leading limit while the second term is the sub-leading limit of the amplitude, which can be compared with the operator statements in \eqref{leadsoft} and \eqref{subsoft} respectively. We have dropped terms of $\order{k_4}$ and higher. We will now try to derive the sub-leading limit of the same correlator ($\langle JOOJ\rangle$) in (EA)dS space by uplifting the amplitude given in \eqref{eq:sqed} via dressing,
\begin{align}\label{eq:JOOJ dressed}
\langle J(\vec{k}_1)O(\vec{k}_2)O(\vec{k}_3)J(\vec{k}_4)\rangle_{(t)}&=V_{14}^{\;i}\pi_{ij}(\vec{k}_2-\vec{k}_3)^j\int \frac{dp}{p^2+|\vec{k}_1+\vec{k}_4|^2}\frac{k_{12}}{p^2+k_{12}^2}\frac{k_{34}}{p^2+k_{34}^2}\notag\\
&+V_{14}^{\;i}\frac{(\vec{k}_1+\vec{k}_4)^i(\vec{k}_1+\vec{k}_4)^j}{|\vec{k}_1+\vec{k}_4|^2}(\vec{k}_2-\vec{k}_3)^j\int \frac{dp}{p^2+|\vec{k}_1+\vec{k}_4|^2}\frac{p}{p^2+k_{12}^2}\frac{p}{p^2+k_{34}^2}.
\end{align}
We first focus on the contribution from the transverse propagator to the correlator in \eqref{eq:JOOJ dressed}. Upon performing the dressing integrals, we obtain
\begin{align}
&\langle J(\vec{k}_1)O(\vec{k}_2)O(\vec{k}_3)J(\vec{k}_4)\rangle_{(t)}^T\notag\\
&=\frac{\big(\epsilon_1\cdot\epsilon_4(k_1-k_4)^i+2(\epsilon_1\cdot k_4)\epsilon_4^i-2(\epsilon_4\cdot k_1)\epsilon_1^i\big)\;(\vec{k}_2-\vec{k}_3)^j}{2(k_1+k_2+k_3+k_4)(k_1+k_2+|\vec{k}_1+\vec{k}_4|)(k_2+k_3+|\vec{k}_1+\vec{k}_4|)}\Bigg(\eta_{ij}-\frac{(\vec{k}_1+\vec{k}_4)_i(\vec{k}_1+\vec{k}_4)_j}{|\vec{k}_1+\vec{k}_4|^2}\Bigg),
\end{align}
which, upon careful application of momentum conservation, can be recast in the following form,
\begin{align}\label{eq:JOOJ sub}
&\langle J(\vec{k}_1)O(\vec{k}_2)O(\vec{k}_3)J(\vec{k}_4)\rangle_{(t)}^T\notag\\
&=\frac{\big(-\epsilon_1\cdot\epsilon_4(k_2+k_3)^i-2\epsilon_1\cdot\epsilon_4 k_4^i+2(\epsilon_1\cdot k_4)\epsilon_4^i-2(\epsilon_4\cdot k_1)\epsilon_1^i\big)\;(\vec{k}_2-\vec{k}_3)^j}{2(k_1+k_2+k_3+k_4)(k_1+k_2+|\vec{k}_1+\vec{k}_4|)(k_2+k_3+|\vec{k}_1+\vec{k}_4|)}\Bigg(\eta_{ij}-\frac{(\vec{k}_1+\vec{k}_4)_i(\vec{k}_1+\vec{k}_4)_j}{|\vec{k}_1+\vec{k}_4|^2}\Bigg).
\end{align}
We now perform a systematic expansion of \eqref{eq:JOOJ sub} in the soft momentum $\vec{k}_4$. In doing so, we carefully track the origin of each contribution, distinguishing between terms arising from the dressing factors and those originating from the flat space amplitude itself (see \eqref{eq:JOOJ dressed}). As we will see, this separation allows for a more transparent organization of the result. Expanding up to $\mathcal{O}(k_4)$ we get the following expression,
\begin{align}\label{eq:JOOJ subleading}
&\lim_{\vec{k}_4\rightarrow0}\langle J(\vec{k}_1)O(\vec{k}_2)O(\vec{k}_3)J(\vec{k}_4)\rangle_{(t)}^T\notag\\
&=\frac{1}{2k_1(k_1+k_2+k_3)^2}\Big(1-\frac{k_4}{2k_1}-\frac{k_4}{k_1+k_2+k_3}-\frac{2\vec{k}_1\cdot\vec{k}_4}{k_1^2}\Big)\Big(\eta_{ij}-\frac{\vec{k}_{1i}\vec{k}_{1j}}{k_1^2}\Big(1-\frac{2\vec{k}_1\cdot\vec{k}_4}{k_1^2}\Big)-\frac{\vec{k}_{1i}\vec{k}_{4j}+\vec{k}_{1j}\vec{k}_{4i}}{k_1^2}\Big)\notag\\
&\quad\big(-\epsilon_1\cdot\epsilon_4(k_2+k_3)^i-2\epsilon_1\cdot\epsilon_4 k_4^i+2(\epsilon_1\cdot k_4)\epsilon_4^i-2(\epsilon_4\cdot k_1)\epsilon_1^i\big)\;(\vec{k}_2-\vec{k}_3)^j,
\end{align}
where we have retained terms only up to $\mathcal{O}(k_4)$, as the subleading contribution in the soft limit is linear in $k_4$. We then separate the contribution arising from the leading and subleading soft limit of the amplitude from the remaining terms in the correlator as follows,
\begin{align}\label{eq:JOOJ sub final}
\lim_{\vec{k}_4\rightarrow0}\langle J(\vec{k}_1)O(\vec{k}_2)O(\vec{k}_3)J(\vec{k}_4)\rangle_{(t)}^T&=\frac{\big(\vec{k}_1 \cdot \vec{\epsilon}_4 \; \vec{k}_2 \cdot \vec{\epsilon}_1
-\, \vec{k}_1 \cdot \vec{\epsilon}_4 \; \vec{k}_3 \cdot \vec{\epsilon}_1\big)}{k_1(k_1+k_2+k_3)}\Big(\frac{k_4}{2k_1}+\frac{k_4}{k_1+k_2+k_3}+\frac{2\vec{k}_1\cdot\vec{k}_4}{k_1^2}\Big)\notag\\
&+\frac{(\vec{k}_2 \cdot \vec{\epsilon}_4 \; \vec{k}_4 \cdot \vec{\epsilon}_1
- \, \vec{k}_3 \cdot \vec{\epsilon}_4 \; \vec{k}_4 \cdot \vec{\epsilon}_1
- \, \vec{k}_2 \cdot \vec{k}_4 \; \vec{\epsilon}_1 \cdot \vec{\epsilon}_4
+ \, \vec{k}_3 \cdot \vec{k}_4 \; \vec{\epsilon}_1 \cdot \vec{\epsilon}_4)}{2\, k_1 \left(k_1 + k_2 + k_3\right)^2}\notag\\
&-\frac{\left( k_2 \cdot k_4 - k_3 \cdot k_4 \right)
(\epsilon_1 \cdot \epsilon_4)
}{
4\, k_1 \left(k_1 + k_2 + k_3\right)^2
}
-
\frac{
3 \left( k_2^2 - k_3^2 \right) \, (k_1 \cdot k_4)\, (\epsilon_1 \cdot \epsilon_4)
}{
4\, k_1^3 \left(k_1 + k_2 + k_3\right)^2
},
\end{align}
where we have used momentum conservation and kinematics for three-point. Thus, we see that the sub-leading soft limit of the correlator given in \eqref{eq:JOOJ subleading} can be arranged in the following manner:
\newline
1. The first line of \eqref{eq:JOOJ sub final} arises from the subleading soft expansion of the dressing, multiplied with the leading soft theorem in flat space, as given in \eqref{tsoft}. Therefore, one can write this piece in the following manner,
\begin{align}\label{leadsqed}
&\Bigg(\frac{1}{k_1(k_1+k_2+k_3)^2}\Big(\frac{k_4}{2k_1}+\frac{k_4}{k_1+k_2+k_3}+\frac{2\vec{k}_1\cdot\vec{k}_4}{k_1^2}\Big)\Bigg)\big(\vec{k}_1 \cdot \vec{\epsilon}_4 \; \vec{k}_2 \cdot \vec{\epsilon}_1
-\, \vec{k}_1 \cdot \vec{\epsilon}_4 \; \vec{k}_3 \cdot \vec{\epsilon}_1\big)\notag\\
&=\left(\frac{k_4}{2 k_1}+ \frac{k_4}{k_1 + k_2 + k_3}
+ \frac{2\, \vec{k}_1 \cdot \vec{k}_4}{k_1^2}\right)\frac{\vec{\epsilon}_4\cdot \vec{k}_1}{2k_1}\Bigg(\frac{\vec{\epsilon}_1\cdot( \vec{k}_2-\vec{k}_3)}{(k_1+k_2+k_3)^2}\Bigg)\notag\\
&=\left(\frac{k_4}{2 k_1}+ \frac{k_4}{k_1 + k_2 + k_3}
+ \frac{2\, \vec{k}_1 \cdot \vec{k}_4}{k_1^2}\right)\frac{\vec{\epsilon}_4\cdot \vec{k}_1}{2k_1}\partial_{k_1}\langle J(\vec{k}_1)O(\vec{k}_2)O(\vec{k}_3)\rangle.   
\end{align}\label{subsqed}
2. The second line of \eqref{eq:JOOJ sub final} arises from the leading-order soft limit of the dressing, multiplied with the subleading soft theorem in flat space, as given in \eqref{tsoft} and can be written as follows,
\begin{align}
&\Bigg(\frac{1}{4\, k_1 \left(k_1 + k_2 + k_3\right)^2}\Bigg)\Big(\vec{k}_2 \cdot \vec{\epsilon}_4 \; \vec{k}_4 \cdot \vec{\epsilon}_1
- \vec{k}_3 \cdot \vec{\epsilon}_4 \; \vec{k}_4 \cdot \vec{\epsilon}_1
- \vec{k}_2 \cdot \vec{k}_4 \; \vec{\epsilon}_1 \cdot \vec{\epsilon}_4
+ \vec{k}_3 \cdot \vec{k}_4 \; \vec{\epsilon}_1 \cdot \vec{\epsilon}_4\Big)\notag\\
&=\frac{\vec{\epsilon}_4^{\;i} \vec{k}_4^{\;j}}{2k_1}\Big(\epsilon_1^i\frac{\partial}{\partial\epsilon_1^j}-\epsilon_1^j\frac{\partial}{\partial\epsilon_1^i}+k_1^i\frac{\partial}{\partial k_1^j}-k_1^j\frac{\partial}{\partial k_1^i}\Big)\Bigg(\frac{\vec{\epsilon}_1\cdot( \vec{k}_2-\vec{k}_3)}{(k_1+k_2+k_3)^2}\Bigg)\notag\\
&=\frac{\vec{\epsilon}_4^{\;i} \vec{k}_4^{\;j}}{2k_1}J_{1}^{ij}\partial_{k_1}\langle J(\vec{k}_1)O(\vec{k}_2)O(\vec{k}_3)\rangle.
\end{align}
3. In addition, there are off-shell contributions in the third line of \eqref{eq:JOOJ sub final}, that take the following form,
\begin{align}
-\frac{\left(  \vec{k}_2 \cdot \vec{k}_4 - \vec{k}_3 \cdot \vec{k}_4 \right)(\vec{\epsilon}_1 \cdot \vec{\epsilon}_4)}
{4\, k_1 \left(k_1 + k_2 + k_3\right)^2}
-
\frac{3 \left( k_2^2 - k_3^2 \right)\,(\vec{k}_1 \cdot \vec{k}_4)\, (\vec{\epsilon}_1 \cdot \vec{\epsilon}_4)}
{4\, k_1^3 \left(k_1 + k_2 + k_3\right)^2}.
\end{align}
Thus, up to the off-shell contributions, we find that the subleading limit of the (EA)dS correlator can be obtained by appropriately uplifting the flat-space soft theorems. 

Having established the general structure of subleading soft limits and their relation to dressed flat-space soft theorems for SQED, we now turn to a concrete example in the context of Yang–Mills theory. In particular, we analyze the subleading soft behavior of the four-point correlator, examining how the interplay between gauge structure and dressing manifests in this setting. Consider the color ordered $\langle JJJJ\rangle$ correlator. The sub-leading soft limit of this correlator can yet again be segregated into three categories as follows, \newline
1. Sub-leading order soft dressing of the leading soft theorem
\begin{align}\label{leadYM}
\left(\frac{k_4}{2 k_1}+ \frac{k_4}{k_1 + k_2 + k_3}
+ \frac{2\, \vec{k}_1 \cdot \vec{k}_4}{k_1^2}\right)\frac{\vec{\epsilon}_4\cdot \vec{k}_1}{2k_1}\partial_{k_1}\langle J(\vec{k}_1,\vec{\epsilon}_1)J(\vec{k}_2,\vec{\epsilon}_2)J(\vec{k}_3,\vec{\epsilon}_3)\rangle\notag\\
-\left(\frac{k_4}{2 k_3}+ \frac{k_4}{k_1 + k_2 + k_3}
+ \frac{2\, \vec{k}_3 \cdot \vec{k}_4}{k_3^2}\right)\frac{\vec{\epsilon}_4\cdot \vec{k}_3}{2k_3}\partial_{k_3}\langle J(\vec{k}_1,\vec{\epsilon}_1)J(\vec{k}_2,\vec{\epsilon}_2)J(\vec{k}_3,\vec{\epsilon}_3)\rangle.   
\end{align}
2. Leading order soft dressing of the sub-leading soft theorem
\begin{align}\label{subYM}
\frac{\vec{\epsilon}_4^{\;i} \vec{k}_4^{\;j}}{2k_1}J_{1}^{ij}\partial_{k_1}\langle J(\vec{k}_1,\vec{\epsilon}_1)J(\vec{k}_2,\vec{\epsilon}_2)J(\vec{k}_3,\vec{\epsilon}_3)\rangle-\frac{\vec{\epsilon}_4^{\;i} \vec{k}_4^{\;j}}{2k_3}J_{3}^{ij}\partial_{k_3}\langle J(\vec{k}_1,\vec{\epsilon}_1)J(\vec{k}_2,\vec{\epsilon}_2)J(\vec{k}_3,\vec{\epsilon}_3)\rangle.
\end{align}
3. Off-shell parts at sub-leading limit: The off-shell parts are in general clumsy and do not have any nice operator form at sub-leading order. We omit the expression to avoid clutter.

Notice that the sub-leading soft expansion in \eqref{leadsqed} for scalar QED and \eqref{leadYM} for Yang-Mills theory are the same. This is because they both have same dressing factor \eqref{eq:phi4dressing}, owing to the fact that spin-1 currents and conformally coupled scalars have same dressing. However, since this factor is theory dependent, the sub-leading expansion of dressing is not universal. Thus, we observe that a significant part of the sub-leading limit of (EA)dS correlator can be uplifted via appropriate dressings of flat-space soft theorems.


\section{Summary and future directions}\label{sec:Discussion}

Motivated by recent progress revealing a close structural connection between de Sitter observables and flat-space scattering amplitudes, we revisit cosmological correlators from the perspective of dressed flat-space physics. In this approach, the complicated time-dependent structure of dS computations is reorganised so that the essential dynamical content is inherited from flat-space amplitudes, while the effects of the expanding background are encoded in auxiliary propagators. This viewpoint not only provides a unifying framework for understanding known results like cosmological cutting rules, but also suggests that a wider class of properties in de Sitter space may admit a similar interpretation. In this work, we explore this idea and demonstrate that several nontrivial features of cosmological correlators can indeed be derived by appropriately uplifting and dressing flat-space structures.

\noindent Below we summarize the main results of the paper:

\begin{itemize}
    \item We derive cosmological cutting rules for spinning correlators in (EA)dS by appropriately dressing the flat-space optical theorem, thereby generalising earlier results, which were restricted to conformally coupled scalar\ \cite{Ansari:2026xkm}, to fields with spin.

    \item We show that the cosmological tree theorem can be understood as a direct consequence of the Feynman tree theorem in flat space, once it is suitably uplifted to the de Sitter setting and dressed with the appropriate time-dependent propagator structure. This provides a transparent origin for the cosmological tree theorem and clarifies its relation to familiar structures in flat-space quantum field theory.

    \item We illustrate that the recursion relation for (EA)dS correlators can simply be obtained by dressing the flat-space BCFW recursion relation.

    \item We demonstrate that leading soft theorems in flat-space gauge theories, when appropriately uplifted to the cosmological setting, naturally reproduce the known soft limits of (EA)dS correlators. This offers a unified perspective on soft limits in (EA)dS and clarifies their connection to well-established flat-space physics.

    \item While subleading soft limits in flat space are known to possess a universal structure, no analogous universality has been established for (EA)dS correlators. We show that, once the appropriate dressing factors are incorporated, a universal-looking form for subleading soft limits in(EA)dS does in fact emerge. By systematically reorganizing these effects through dressing, we recover a universal behaviour that closely mirrors its flat-space counterpart, thereby extending the correspondence between flat-space amplitudes and cosmological observables beyond the leading soft limit.

\end{itemize}

More broadly, our results provide strong evidence that a wide range of structural properties of cosmological correlators, including factorization, soft behavior, and discontinuity relations, can be systematically understood as consequences of flat-space physics once appropriately dressed for a curved background. This perspective not only unifies several previously disparate observations, but also suggests a general organizing principle for cosmological observables, in which their analytic structure and consistency conditions are inherited from, and closely tied to, their flat-space counterparts. There are, however,  some interesting explorations that we leave for future work.

\vspace{0.2cm}

\noindent\textbf{Positivity constraints from dressing}

\vspace{0.2cm}
\noindent It would be interesting to explore whether positivity bounds in de Sitter space can be systematically derived by dressing their flat-space counterparts. In particular, one may ask if the well-established positivity constraints on scattering amplitudes translate into analogous constraints on cosmological correlators once the appropriate time-dependent dressing is incorporated.

\vspace{0.2cm}

\noindent\textbf{Beyond perturbation theory}

\vspace{0.2cm}
\noindent An important open question is whether the dressing prescription can be extended beyond the perturbative regime. It would be valuable to investigate if a non-perturbative formulation exists in which the full wavefunction of the universe, or exact correlators, can be expressed in terms of suitably dressed flat-space data. Such an extension could provide new insights into resummation of perturbative series, non-perturbative effects, and the underlying structure of quantum field theory in cosmological spacetimes.

\vspace{0.2cm}

\noindent\textbf{Connection to flat space holography}

\vspace{0.2cm}
\noindent It would be interesting to explore whether the dressing framework provides a bridge to flat-space holography, including structures such as Carroll symmetry and celestial holography. In particular, one may ask whether the dressed representation of cosmological correlators makes these symmetries more manifest, and if it allows for a unified description in which cosmological observables can be mapped to celestial amplitudes or Carrollian field theories.

\vspace{0.2cm}

\noindent\textbf{Bootstrap via dressing}

\vspace{0.2cm}
\noindent An interesting direction is to leverage the dressing framework to import S-matrix bootstrap techniques into the cosmological setting. In particular, one may attempt to construct cosmological correlators by dressing flat-space amplitudes that satisfy consistency conditions such as unitarity, analyticity, and locality, thereby formulating a bootstrap program for cosmological observables grounded in well-established flat-space principles.
\acknowledgments
We thank Sachin Jain for useful discussions during the course of this work. We acknowledge our debt to the people of India for their steady support of research in basic sciences.


\appendix
\section{Notations and Conventions}\label{app:note}
In this appendix, we present all the notations and conventions used throughout the paper.
\subsection*{Notations}
The four-dimensional spacetime vector indices are labeled with the Greek alphabets, e.g. ${\mu}\in\{0,1,2,3\}$, with the signature of metric given by $\eta_{{\mu}\nu}=\textbf{diag}(\mp,-,-,-)$ for Euclidean/Lorentzian case, where zeroth component is the axial/temporal respectively. Meanwhile, the boundary directions are labeled with lowercase alphabets e.g. ${i}\in\{1,2,3\}$.

\noindent The contraction of vectors are denoted by a dot product defined as
\begin{align}
A^\mu B_\mu=A\cdot B,\notag\\
A^{\;i}b_{\;i}=\vec{A}\cdot\vec{B}.
\end{align}
We use further shorthand notations, based on \cite{Chowdhury:2024wwe} for convenience
\begin{align}
k_i:&\qquad\textrm{Energy of $i^\textrm{th}$ external leg.}\notag\\
p_i:&\qquad\textrm{Energy of $i^\textrm{th}$ internal non-auxiliary propagator.}\notag\\
k_{ij\cdots}:&\qquad |\vec{k}_i|+|\vec{k}_j|+\cdots\notag\\
\vec{y}_{ij\cdots}:&\qquad \vec{k}_i+\vec{k}_j+\cdots
\end{align}
\subsection*{Conventions}
We work in the Poincare patch of dS$_4$, where the metric is given by
\begin{align}\label{metricdS}
ds^2=R_\textrm{dS}^2\frac{-d\eta^2+d\vec{x}^2}{\eta^2},
\end{align}
where $R_\textrm{dS}$ denotes the radius of curvature, $\eta\in\{-\infty,0\}$ is the conformal time, and $\vec{x}$ denotes the Euclidean boundary directions.

The (EA)dS$_4$ setup is then obtained by a double Wick-rotation, wherein $\eta\rightarrow z=-i\eta$ and $R_\textrm{dS}\rightarrow R_\textrm{EAdS}=iR_\textrm{dS}$. The corresponding metric is given by
\begin{align}\label{metricEAdS}
ds^2=R_\textrm{EAdS}^2\frac{dz^2+d\vec{x}^2}{z^2}.
\end{align}
We will set $R_\textrm{EAdS}=1$ for convenience.

We use the following definitions for discontinuity in this work
\begin{equation}\label{eq:Disc}
    \mathrm{Disc}_s f(s^2) = f(s^2+i\epsilon)-f(s^2-i\epsilon),
\end{equation}
\begin{equation}\label{eq:disc}
    \mathrm{disc}_s f(s) = \frac{1}{2}\left(f(s) - f(-s)\right),
\end{equation}
\begin{equation}\label{eq:discbar}
    \overline{\mathrm{disc}}_s f(s) = \frac{1}{2}\left(f(s) + f(-s)\right).
\end{equation}

In all the scattering amplitude calculations, we will follow the convention where all momenta are outgoing.

\section{Cutting Rules for In-In Correlators}\label{app:cutinin}

In section \ref{sec:Cutting}, we discussed the cosmological cutting rules for the wavefunction coefficients involving spinning fields. In this appendix, we will present the results of cutting rules for in-in correlators in the spinning case.

\subsection*{Tree-level}
As an illustrative example, consider the s-channel tree-level exchange diagram for scalar QED. The corresponding amplitude is given as ,
\begin{align}
\mathcal{A}_{4,s}^\textrm{Tree}=-g^2\frac{\alpha^\mu\pi_{\mu\nu}\beta^\nu}{(k_1+k_2)^2},
\end{align}
where $\alpha^\mu=(k_1-k_2)^\mu\;\textrm{and}\;\beta^\nu=(k_3-k_4)^\nu$. The amplitude is uplifted to its corresponding in-in correlator by a similar dressing procedure as discussed in section \ref{sec:Cutting}
\begin{align}\label{ininsqed}
\langle \phi(\vec{k}_1)\phi^\dagger(\vec{k}_2) \phi(\vec{k}_3) \phi^\dagger(\vec{k}_4) \rangle_{(s)}&=\;\alpha^i\pi_{ij}\beta^j\int_{-\infty}^{\infty} dp \Delta_{\gamma}^T(k_{12},p)\Delta_{\gamma}^T(k_{34},p)\frac{1}{p^2+|\vec{s}|^2}\notag\\
&+\;\alpha^i\beta^j\frac{s_is_j}{|\vec{s}|^2}\int_{-\infty}^{\infty} dp \Delta_{\gamma}^L(k_{12},p)\Delta_{\gamma}^L(k_{34},p)\frac{1}{p^2},
\end{align}
where the dressing factors for this amplitude are as follows \cite{Chowdhury:2025nnk}
\begin{align}\label{ininsqeddress}
\Delta_{\gamma}^T(k_{ext},p)&=\int dz\; e^{-k_{ext}z}\;\textrm{cos}(pz)=\frac{k_{ext}}{p^2+k_{ext}^2},\\
\Delta_{\gamma}^L(k_{ext},p)&=\int dz\; e^{-k_{ext}z}\;\textrm{sin}(pz)=\frac{p}{p^2+k_{ext}^2}.
\end{align}
Recall the statement of flat-space optical theorem for this case in \eqref{eq:opticalthm2p}
\begin{align}\label{QEDoptthmin}
2\,\mathrm{Im}\,\mathcal{A}_{4,s}^{\text{Tree}}
= \sum_h \int \frac{d^3 q}{(2\pi)^3}\,\frac{1}{2|\vec{q}|}\,\mathcal{A}_3(k_1 k_2 \rightarrow q)\mathcal{A}^{*}_3(k_3 k_4 \rightarrow q)\,
(2\pi)^4\delta^{4}(k_1 + k_2 - q).
\end{align}
After dressing the above equation appropriately with \eqref{ininsqeddress}, the RHS of \eqref{QEDoptthmin} becomes
\begin{align}
&\sum_h  \frac{1}{2|\vec{s}|}\left( g\,\alpha^{\;i}\epsilon^h_i(\vec{s}) 
\int_{0}^{\infty} dz_1\, e^{-k_{12} z_1}\,\mathrm{cosh}(s\, z_1)\right)\,\left(g\, \beta^{\;j}  \epsilon^{*\,h}_j(\Vec{s}) \,\int_{0}^{\infty} dz_2\, e^{-k_{34} z_2}\,\mathrm{cosh}(s\,z_2)\right)\nonumber \\
& = \sum_h \frac{1}{2|\vec{s}|} \left(\overline{\mathrm{disc}}_s  \langle \phi(\vec{k}_1) \phi^\dagger(\vec{k}_2)\phi(\vec{s })\rangle^h\right)\,\left(\overline{\mathrm{disc}}_s  \langle \phi(-\vec{s}) \phi(\vec{k}_3)\phi^\dagger(\vec{k}_4) \rangle^h\right),
\end{align}
Doing the same, but now for the LHS of \eqref{QEDoptthmin} yields
\begin{align}
&g^{2}\,\alpha^i\pi_{ij}\beta^j\, \int_{-\infty}^{\infty}
dp\frac{k_{12}}{p^2+k_{12}^2}\frac{k_{34}}{p^2+k_{34}^2}(2\pi)\delta\!\left(p^{2}+s^{2}\right)+
g^{2}\,\alpha^i\beta^j\frac{s_is_j}{|\vec{s}|^2}\, \int_{-\infty}^{\infty}
dp\frac{p}{p^2+k_{12}^2}\frac{p}{p^2+k_{12}^2}\delta\!\left(p^{2}\right)\notag\\
&=i \, \mathrm{Disc}_s \left(\langle \phi(\vec{k}_1)\phi^\dagger(\vec{k}_2) \phi(\vec{k}_3) \phi^\dagger(\vec{k}_4) \rangle_{(s)}\right),
\end{align}
which is obtained by sending the internal propagator on-shell for both cases. Notice the second term drops off, implying that the action of discontinuity in the contribution from the longitudinal part of the propagator is zero. This leads to
\begin{align}\label{ininLHS}
i \, \mathrm{Disc}_s \left(\langle \phi(\vec{k}_1)\phi^\dagger(\vec{k}_2) \phi(\vec{k}_3) \phi^\dagger(\vec{k}_4) \rangle_{(s)}\right)&=g^{2}\,\alpha^i\pi_{ij}\beta^j\, \int_{-\infty}^{\infty}
dp\frac{k_{12}}{p^2+k_{12}^2}\frac{k_{34}}{p^2+k_{34}^2}\delta\!\left(p^{2}+s^{2}\right)\notag\\
&=g^{2}\sum_h\alpha^i\beta^j\epsilon^{h}_j(\Vec{s})\epsilon^{*\,h}_j(\Vec{s})\, \int_{-\infty}^{\infty}
dp\frac{k_{12}}{p^2+k_{12}^2}\frac{k_{34}}{p^2+k_{34}^2}\delta\!\left(p^{2}+s^{2}\right)
\end{align}
where we have used the completeness relation \eqref{eq:completeness}. After performing the energy integral in \eqref{ininLHS}, we obtain the cutting rule for in-in correlator
\begin{equation}
    \mathrm{Disc}_s \left(\langle \phi(\vec{k}_1)\phi^\dagger(\vec{k}_2) \phi(\vec{k}_3) \phi^\dagger(\vec{k}_4) \rangle\right)_{(s)} = -i \sum_h \frac{1}{2\abs{\vec{s}}} \left(\overline{\mathrm{disc}}_s  \langle \phi(\vec{k}_1) \phi^\dagger(\vec{k}_2)\phi(\vec{s })\rangle^h_L\right)\,\left(\overline{\mathrm{disc}}_s  \langle \phi(-\vec{s}) \phi(\vec{k}_3)\phi^\dagger(\vec{k}_4) \rangle^h_R\right).
\end{equation}
\subsection*{Loops}
We now move to the case of 1-loop four-point function in s-channel. The statement of optical theorem for this diagram is given in \eqref{eq:imA4loop}. Separately, the imaginary part of the amplitude is as follows \eqref{eq:lhsoptSQED4p1l}
\begin{align}\label{lhsoptSQED4p1lin}
    2\mathrm{Im}(\mathcal{A}_4^{1-loop}) &= g^4\,\int d^3 l_1\,d^3 l_2\, dp_1\, dp_2 \, \pi_{jl}(\vec{l}_1)\,\pi_{jl}(\vec{l}_2)\,(2\pi)^2\delta(p_1^2- l_1^2)\,\delta(p_2^2-l_2^2)\,(2\pi)^4\delta^4(K_1+K_2+L_1-L_2).
\end{align}
The shadow action for scalar QED allows for both: exchange of the Dirichlet (++) and the Neumann ($--$) mode for both the internal photons\footnote{Please refer equation 3.19 of \cite{Chowdhury:2025nnk} for the relevant shadow action.}. Starting from \eqref{lhsoptSQED4p1lin}, and dressing with the ($--$) exchange of both transverse and longitudinal parts of photon propagator, leads to 
\begin{align}\label{LHS--}
&  \mathrm{Disc}_{l_1} \mathrm{Disc}_{l_2} (\langle  \phi_1(\vec{k}_1)\phi_2(\vec{k}_2)\phi_3(\vec{k}_3)\phi_4(\vec{k}_4)\rangle^{--})\notag\\
&=\pi_{jl}\,\pi_{jl}\, \int\, dp_1\, dp_2\, dz_1\, dz_2\, e^{-k_{12} z_1}\, \mathrm{cos}(p_1 z_1)\,  \mathrm{cos}(p_1 z_2)\,e^{-k_{34} z_2}\, \mathrm{cos}(p_2 z_1)\,  \mathrm{cos}(p_2 z_2)\,  \delta(p_1^2+ l_1^2)\,\delta(p_2^2+l_2^2).
\end{align}
Doing the same for (++) exchange results in
\begin{align}\label{LHS++}
   &  \mathrm{Disc}_{l_1} \mathrm{Disc}_{l_2} (\langle  \phi_1(\vec{k}_1)\phi_2(\vec{k}_2)\phi_3(\vec{k}_3)\phi_4(\vec{k}_4)\rangle^{++})\notag \\
   & = \pi_{jl}\,\pi_{jl}\, \int\, dp_1\, dp_2\, dz_1\, dz_2\, e^{-k_{12} z_1}\, \mathrm{sin}(p_1 z_1)\,  \mathrm{sin}(p_1 z_2)\,e^{-k_{34} z_2}\, \mathrm{sin}(p_2 z_1)\,  \mathrm{sin}(p_2 z_2)\,  \delta(p_1^2+ l_1^2)\,\delta(p_2^2+l_2^2).
\end{align}
Thus, the total action of discontinuity on \eqref{lhsoptSQED4p1lin} comes by combining \eqref{LHS--} and \eqref{LHS++}
\begin{align}\label{LHS}
& \mathrm{Disc}_{l_1} \mathrm{Disc}_{l_2} (\langle  \phi_1(\vec{k}_1)\phi_2(\vec{k}_2)\phi_3(\vec{k}_3)\phi_4(\vec{k}_4)\rangle)\notag\\
&=\pi_{jl}\,\pi_{jl}\, \int\, dp_1\, dp_2\, dz_1\, dz_2\, e^{-k_{12} z_1}\, \mathrm{cos}(p_1 z_1)\,  \mathrm{cos}(p_1 z_2)\,e^{-k_{34} z_2}\, \mathrm{cos}(p_2 z_1)\,  \mathrm{cos}(p_2 z_2)\,  \delta(p_1^2+ l_1^2)\,\delta(p_2^2+l_2^2)\notag \\
&+\pi_{jl}\,\pi_{jl}\, \int\, dp_1\, dp_2\, dz_1\, dz_2\, e^{-k_{12} z_1}\, \mathrm{sin}(p_1 z_1)\,  \mathrm{sin}(p_1 z_2)\,e^{-k_{34} z_2}\, \mathrm{sin}(p_2 z_1)\,  \mathrm{sin}(p_2 z_2)\,  \delta(p_1^2+ l_1^2)\,\delta(p_2^2+l_2^2).
\end{align}
We now turn our focus to the RHS of \eqref{eq:imA4loop}. Dressing for ($--$) dressing gives
\begin{align}\label{RHS--}
     & \pi_{jl}\,\pi_{jl}\, \int\, \frac{dp_1}{2p_1}\,\frac{dp_2}{2p_2}\,dz_1\,dz_2\, e^{-k_{12} z_1}\, \mathrm{cos}(p_1 z_1)\,  \mathrm{cos}(p_1 z_2)\,e^{-k_{34} z_2}\, \mathrm{cos}(p_2 z_1)\,  \mathrm{cos}(p_2 z_2)\,  \delta(p_1 + i l_1)\,\delta(p_2+i l_2)\nonumber \\
     & = \sum_{hh'} \frac{1}{4|\vec{l}_1||\vec{l}_2|} \left(\overline{\mathrm{disc}}_{l_1} \overline{\mathrm{disc}}_{l_2} (\langle  \phi_1(\vec{k}_1)\phi_2(\vec{k}_2)\phi_3(\vec{l}_1)\phi_4(\vec{l}_2)\rangle)\right)  \left(\overline{\mathrm{disc}}_{l_1} \overline{\mathrm{disc}}_{l_2} (\langle  \phi_1(-\vec{l}_1)\phi_2(-\vec{l}_2)\phi_3(\vec{k}_3)\phi_4(\vec{k}_4)\rangle)\right), 
\end{align}
while doing the same for (++) dressing leads to
\begin{align}\label{RHS++}
     & \pi_{jl}\,\pi_{jl}\, \int\, \frac{dp_1}{2p_1}\,\frac{dp_2}{2p_2}\,dz_1\,dz_2\, e^{-k_{12} z_1}\, \mathrm{sin}(p_1 z_1)\,  \mathrm{sin}(p_1 z_2)\,e^{-k_{34} z_2}\, \mathrm{sin}(p_2 z_1)\,  \mathrm{sin}(p_2 z_2)\,  \delta(p_1 - i l_1)\,\delta(p_2-i l_2)\nonumber \\
     & =  \sum_{hh'}\frac{1}{4|\vec{l}_1||\vec{l}_2|} \left(\mathrm{disc}_{l_1} \mathrm{disc}_{l_2} (\langle  \phi_1(\vec{k}_2)\phi_2(\vec{k}_2)\phi_3(\vec{l}_1)\phi_4(\vec{l}_2)\rangle)\right)  \left(\mathrm{disc}_{l_1} \mathrm{disc}_{l_2} (\langle  \phi_1(-\vec{l}_1)\phi_2(-\vec{l}_2)\phi_3(\vec{k}_3)\phi_4(\vec{k}_4)\rangle)\right). 
\end{align}
Using optical theorem to compare the LHS \eqref{LHS} and RHS \eqref{RHS--} together with \eqref{RHS++}, we arrive at following relation,
\begin{align}
&\mathrm{Disc}_{l_1} \mathrm{Disc}_{l_2} (\langle  \phi_1(\vec{k}_1)\phi_2(\vec{k}_2)\phi_3(\vec{k}_3)\phi_4(\vec{k}_4)\rangle)\nonumber \\
&=\frac{1}{4|\vec{l}_1||\vec{l}_2|} \left(\mathrm{disc}_{l_1} \mathrm{disc}_{l_2} (\langle  \phi_1(\vec{k}_1)\phi_2(\vec{k}_2)\phi_3(\vec{l}_1)\phi_4(\vec{l}_2)\rangle)\right)  \left(\mathrm{disc}_{l_1} \mathrm{disc}_{l_2} (\langle  \phi_1(-\vec{l}_1)\phi_2(-\vec{l}_2)\phi_3(\vec{k}_3)\phi_4(\vec{k}_4)\rangle)\right)\notag\\
&+\frac{1}{4|\vec{l}_1||\vec{l}_2|} \left(\overline{\mathrm{disc}}_{l_1} \overline{\mathrm{disc}}_{l_2} (\langle  \phi_1(\vec{k}_1)\phi_2(\vec{k}_2)\phi_3(\vec{l}_1)\phi_4(\vec{l}_2)\rangle)\right)  \left(\overline{\mathrm{disc}}_{l_1} \overline{\mathrm{disc}}_{l_2} (\langle  \phi_1(-\vec{l}_1)\phi_2(-\vec{l}_2)\phi_3(\vec{k}_3)\phi_4(\vec{k}_4)\rangle)\right).
\end{align}
A similar story holds for correlators in Yang-Mills theory, minimally coupled scalars and Einstein gravity theory as well, but we do not present it here.


\section{Dressing flat space propagators}\label{app:dressprop}

In this section, we explicitly construct the propagators obtained by uplifting their flat-space counterparts through an appropriate dressing procedure.

\subsection*{Feynman propagator}

Starting from the flat space Feynman propagator,
\begin{equation}
    \Pi_F(p,s) =  \frac{i}{p^2-s^2+i\epsilon }
\end{equation}
Applying the appropriate dressing factors as in \eqref{eq:phi4dressing} , we obtain,
\begin{align}
    \Pi^d_F(z_1,z_2,s) &=\int dp\, \frac{i}{p^2-s^2+i\epsilon }\, \mathrm{sin}(p z_1)\, \mathrm{sin}(p z_2)\nonumber \\ 
    & = -\frac{\pi}{2s}\left[e^{-is(z_1+z_2)}- e^{-is(z_1-z_2)}\,\theta(z_1-z_2)-e^{-is(z_2-z_1)}\,\theta(z_2-z_1)\right]
\end{align}
which, upon performing the Wick rotation $s \rightarrow -i s$, matches the Feynman propagator in(EA)dS \eqref{eq:btb}.

\subsection*{Retarded propagator}
\noindent We start with the flat-space retarded propagator defined as,

\begin{equation}\label{feyanti}
    \Pi_R = \frac{1}{2}\left(\Pi_F + \Pi_{\Tilde{F}}- \frac{ \pi}{s}(\delta(p+s)- \delta(p-s))\right)
\end{equation}
dressed Feynman propagator is given as,
\begin{align}
    \Pi^d_F(z_1,z_2,s) &=\int dp\, \frac{i}{p^2-s^2+i\epsilon }\, \mathrm{sin}(p z_1)\, \mathrm{sin}(p z_2)\nonumber \\ 
    & = -\frac{\pi}{2s}\left[e^{-is(z_1+z_2)}- e^{-is(z_1-z_2)}\,\theta(z_1-z_2)-e^{-is(z_2-z_1)}\,\theta(z_2-z_1)\right]
\end{align}
dressed anti-Feynman propagator is given as,
\begin{align}
    \Pi^d_{\tilde{F}}(z_1,z_2,s) &=\int dp\, \frac{i}{p^2-s^2+i\epsilon }\, \mathrm{sin}(p z_1)\, \mathrm{sin}(p z_2)\nonumber \\ 
    & = -\frac{\pi}{2s}\left[-e^{is(z_1+z_2)}+ e^{is(z_1-z_2)}\,\theta(z_1-z_2)+e^{is(z_2-z_1)}\,\theta(z_2-z_1)\right]
\end{align}
with these we can write dressed retarded propagator as,
\begin{align}\label{eq:dressedpir}
     \Pi^d_R(z_1,z_2,s) &=-\frac{\pi}{4 s}\Bigg[\left(e^{- i s (z_1+z_2)}-e^{ i s (z_1+z_2)}\right) - \left(e^{- i s (z_1-z_2)} -e^{ i s (z_1-z_2)}\right)\theta(z_1-z_2)\nonumber \\
    &\qquad \qquad \qquad \qquad \qquad \qquad\quad-\left(e^{- i s (z_2-z_1)} -e^{ i s (z_2-z_1)}\right)\theta(z_2-z_1)\Bigg]
\end{align}
Note that the contributions from the last two $\delta$-functions  in \eqref{feyanti} cancel. We note that the dressed retarded propagator is not exactly the retarded propagator given in \eqref{eq:GREAdS} in(EA)dS. Therefore, we observe that there is no one-to-one correspondence between \eqref{eq:GRdecompoflat} and \eqref{eq:GRdecompoEAdS}.

\section{Dressing four-point 1-loop}\label{ap:calc}

Here, we dress \eqref{eq:A41loop} with the appropriate dressing factors introduced in \eqref{eq:phi4dressing}. The integrand in \eqref{eq:A41loop} consists of four distinct contributions, shown explicitly in \eqref{eq:PiPi}. We now analyze these contributions separately and dress them term by term. This decomposition makes the structure of the dressed loop expression transparent and allows us to identify how each propagator component is modified individually.

\subsection*{$\Pi_{R_1}\Pi_{R_2}$-term}

We first consider the contribution in which both factors are retarded propagators. 

\begin{align}\label{eq:pir1pir2}
    &= \int dz_1 \,dz_2\,\int dp_1\,\,\Pi_R(p_1,q_1)\, \mathrm{sin}(p_1 z_1)\,\mathrm{sin}(p_1 z_2)\int dp_2\,\Pi_R(p_2,q_2)\, \mathrm{sin}(p_2 z_1)\,\mathrm{sin}(p_2 z_2)\nonumber \\
    & = \int dz_1\, dz_2\, \Pi^d_R(z_1,z_2,q_1)\, \Pi^d_R(z_1,z_2,q_2)
\end{align}

where $\Pi^d_R$ is the dressed retarded propagator defined in \eqref{eq:dressedpir}.

\subsection*{$\Pi_{R_1}\delta_2$-term}

Next, we consider the mixed contribution from retarded propagator and $\delta$-function term,

\begin{align}
    &= \int dz_1\,dz_2\,\int dp_1\,\Pi_R(p_1,q_1)\, \mathrm{sin}(p_1 z_1)\,\mathrm{sin}(p_1 z_2)\int dp_2\,\left(\frac{\pi}{q_2}\delta(p_2+q_2)\right)\, \mathrm{sin}(p_2 z_1)\,\mathrm{sin}(p_2 z_2)\nonumber \\
    & = \int dz_1\, dz_2\, \Pi^d_R(z_1,z_2,q_1)\, \left(\frac{\pi}{q_2} \mathrm{sin}(q_2 z_1)\,\mathrm{sin}(q_2 z_2)\right)
\end{align}

\subsection*{$\Pi_{R_2}\delta_1$-term}

The third contribution is completely analogous, with momenta interchanged.

\begin{align}
    &= \int dz_1\,dz_2\,\int dp_2\,\Pi_R(p_2,q_2)\, \mathrm{sin}(p_2 z_1)\,\mathrm{sin}(p_2 z_2)\int dp_1\,\left(\frac{\pi}{q_1}\delta(p_1+q_1)\right)\, \mathrm{sin}(p_1 z_1)\,\mathrm{sin}(p_1 z_2)\nonumber \\
    & = \int dz_1\, dz_2\, \Pi^d_R(z_1,z_2,q_2)\, \left(\frac{\pi}{q_1} \mathrm{sin}(q_1 z_1)\,\mathrm{sin}(q_1 z_2)\right)
\end{align}

\subsection*{$\delta_1\delta_2$-term}

Finally, we consider the purely delta-function contribution. 

\begin{align}\label{eq:d1d2}
    &= \int dz_1\,dz_2\,\int dp_1\,\left(\frac{\pi}{q_1}\delta(p_1+q_1)\right)\, \mathrm{sin}(p_1 z_1)\,\mathrm{sin}(p_1 z_2)\int dp_2\,\left(\frac{\pi}{q_2}\delta(p_2+q_2)\right)\, \mathrm{sin}(p_2 z_1)\,\mathrm{sin}(p_2 z_2)\nonumber \\
    & = \int dz_1\, dz_2\, \left(\frac{\pi}{q_1} \mathrm{sin}(q_1 z_1)\,\mathrm{sin}(q_1 z_2)\right)\left(\frac{\pi}{q_2} \mathrm{sin}(q_2 z_1)\,\mathrm{sin}(q_2 z_2)\right)
\end{align}

Collecting all four contributions, we then perform the Wick rotation $q_i \rightarrow -i q_i$ on the expressions from \eqref{eq:pir1pir2} through \eqref{eq:d1d2}. Summing the resulting terms finally reproduces the expression given in \eqref{eq:GRGREAdS}.

\bibliographystyle{JHEP}
\bibliography{biblio}

\end{document}